\documentclass[12pt,preprint]{emulateapj}

\slugcomment{AJ}

\shorttitle{Vibrationally excited HCN J=4--3 line in IRAS 20551-4250}
\shortauthors{Imanishi et al.}

\begin{document}


\title{ALMA Detection of the Vibrationally Excited HCN J=4--3 Emission
Line in the AGN-Hosting Luminous Infrared Galaxy IRAS 20551$-$4250}


\author{Masatoshi Imanishi\altaffilmark{1,2}}
\affil{Subaru Telescope, 650 North A'ohoku Place, Hilo, Hawaii, 96720,
U.S.A.} 
\email{masa.imanishi@nao.ac.jp}

\and

\author{Kouichiro Nakanishi\altaffilmark{1,2}}
\affil{Joint ALMA Observatory, Alonso de C\'{o}rdova 3107, Vitacura 
763-0355, Santiago de Chile}

\altaffiltext{1}{National Astronomical Observatory of Japan, 2-21-1
Osawa, Mitaka, Tokyo 181-8588}
\altaffiltext{2}{Department of Astronomy, School of Science, Graduate
University for Advanced Studies (SOKENDAI), Mitaka, Tokyo 181-8588}

\begin{abstract}
We present results from our ALMA Cycle 0 observations, at the frequencies 
around the HCN, HCO$^{+}$, and HNC J=4--3 transition lines, of the 
luminous infrared galaxy IRAS 20551$-$4250 at z=0.043, which is known to 
host an energetically important obscured AGN. 
In addition to the targeted HCN, HCO$^{+}$, and HNC J=4--3 emission
lines, two additional strong emission lines are seen, which we
attribute to H$_{2}$S and CH$_{3}$CN$(+$CCH).   
The HCN-to-HCO$^{+}$ J=4--3 flux ratio ($\sim$0.7) is higher than 
in the other starburst-dominated galaxy ($\sim$0.2) observed 
in our ALMA Cycle 0 program. 
We tentatively ($\sim$5$\sigma$) detected the vibrationally 
excited (v$_{2}$=1) HCN J=4--3 (l=1f) 
emission line, which is important for testing an infrared radiative pumping 
scenario for HCN.
This is the second detection of this molecular transition in external
galaxies.  
The most likely reason for this detection is not only the high
flux of this emission line, but also the small molecular line widths
observed in this galaxy, suggesting that vibrational excitation of HCN
may be relatively common in AGN-hosting galaxies.  
\end{abstract}


\section{Introduction}

Luminous infrared galaxies (LIRGs) radiate the bulk of their large
luminosities (L$> 10^{11}$L$_{\odot}$) as infrared dust emission
\citep{sam96}. 
Their large infrared luminosities indicate that powerful energy
sources are present, hidden behind dust, which absorbs most of the
primary energetic radiation, and the heated dust grains reemit this
energy as infrared thermal radiation. 
Since LIRGs are the dominant population at
$z >$ 1, in terms of the cosmic infrared radiation density 
\citep{cap07,got10,mag11,mur11}, determining whether starbursts (i.e.,
nuclear fusion inside stars) are dominant, or whether active galactic
nuclei (AGNs; active mass accretion onto a central compact supermassive
black hole [SMBH] with $>$10$^{6}$M$_{\odot}$) are also energetically
important as obscured energy sources of LIRGs, is closely related to
unveiling the history of star formation and SMBH growth in the
dust-obscured galaxy population of the early universe.      

Unlike optically identifiable AGNs, in which the central mass-accreting
SMBHs are surrounded by a toroidally distributed (torus-shaped) dusty
medium, LIRGs, particularly those with L$_{\rm IR}$ $>$ 10$^{11.5}$L$_{\odot}$
are major mergers of gas-rich galaxies \citep{san04,alo09} and have a large
amount of concentrated molecular gas and dust in their nuclei \citep{sam96}. 
The putative compact AGNs in LIRG nuclei can be easily obscured by gas
and dust in virtually all directions.  
Thus, optical detection of AGN signatures in LIRGs becomes very difficult.
However, evaluating the energetic importance of such optically elusive
{\it buried} AGNs is crucial to understanding the true nature
of the LIRG population \citep{hop06}. 

To investigate buried AGNs in dusty LIRG nuclei, we must observe at
wavelengths in which dust extinction effects are small. 
The (sub)millimeter band is one such set of wavelengths, and the flux
ratios of (sub)millimeter molecular lines for AGNs have been suggested
to be different from the ratios in starburst galaxies
\citep{tac94,koh05,ima04,ima06,in06,ima07,per07,kri08,ima09,gar10,ima10b,cos11}.
For example, an enhancement of the HCN rotational (J) transition line
flux, relative to other molecular lines, is argued in AGNs
\citep{koh05,ima04,ima06,in06,ima07,kri08,ima09}.  
If this is confirmed to be a general trend in AGNs, then (sub)millimeter HCN
emission lines can be used to identify deeply buried AGNs in dusty 
LIRG nuclei. 
Even if strong HCN emission is a feature of AGNs, the interpretation is
still not well established.  
HCN abundance enhancement \citep{lin06,har10} could be a
possible reason for strong HCN emission in AGNs.
Infrared radiative pumping of HCN has also been proposed as
another possible mechanism for the increasing HCN emission flux
\citep{aal95,gar06,gra06,wei07}.  
Namely, an AGN, powered by a mass-accreting SMBH, has
much higher radiative energy generation efficiency than the nuclear
fusion reaction inside stars, can heat the surrounding dust to high
temperatures of several hundred Kelvin, and can emit a much stronger
mid-infrared 3--20 $\mu$m continuum (relative to the bolometric
luminosity) than a starburst.  
By absorbing the infrared 14 $\mu$m photons, HCN
molecules can be vibrationally excited to the v$_{2}$=1 level.
Through subsequent decay to the vibrational ground level (v=0),
the HCN rotational (J) transition lines at v=0 will be strong compared
to collisional excitation alone \citep{ran11}.  
To determine whether this infrared radiative pumping scenario is
generally and efficiently at work in AGNs, the detection of
vibrationally excited HCN J-transition lines in AGN-hosting galaxies is
of particular importance, because the lines are expected to be strong 
in infrared radiative pumping, but not in collisional excitation.

To investigate whether HCN emission is indeed enhanced  
(relative to other molecular lines) in AGNs
compared to starbursts, and to test whether vibrationally excited HCN
emission lines are generally present in AGNs, we observed the HCN,
HCO$^{+}$, and HNC J=4--3 lines of nearby (z$<$0.35) luminous infrared
galaxies, with and without luminous AGN signatures, in ALMA Cycle 0.     
Among six observed sources, the vibrationally excited 
HCN v$_{2}$=1 J=4--3 (l=1f) emission line is tentatively detected 
in the luminous infrared galaxy IRAS 20551$-$4250 at z=0.043, which is 
diagnosed to contain a luminous buried AGN (Table 1).
This is the second detection of this important line in extragalactic sources.
We report this result in this paper.  
Throughout this paper, we adopt H$_{0}$ $=$ 71 km s$^{-1}$ Mpc$^{-1}$, 
$\Omega_{\rm M}$ = 0.27, and $\Omega_{\rm \Lambda}$ = 0.73 \citep{kom09}, 

\section{Observations and Data Analysis}

Our ALMA Cycle 0 observations were performed within the program 
2011.0.00020.S (PI = M. Imanishi).  
Table 2 provides the details of our observations.
The widest 1.875 GHz band mode and 3840 total channel number were 
adopted. 

At the redshift of IRAS 20551$-$4250 (z=0.043), HCN J=4--3 
(rest-frame frequency is $\nu_{\rm rest}$ = 354.505 GHz) and HCO$^{+}$
J=4--3 ($\nu_{\rm rest}$ = 356.734 GHz) lines can be simultaneously
observed in ALMA band 7. 
We employed four spectral windows, of which two were used to cover 
HCN J=4--3 (central frequency was set as  
$\nu_{\rm center}$= 339.890 GHz) and HCO$^{+}$ J=4--3 lines 
($\nu_{\rm center}$ = 341.835 GHz), and the remaining two were 
used to measure the continuum flux level 
($\nu_{\rm center}$ = 351.883 GHz and 353.589 GHz).  

To observe the HNC J=4--3 line ($\nu_{\rm rest}$ = 362.630 GHz), we
needed additional independent observations. 
We covered the HNC J=4--3 line in one spectral window ($\nu_{\rm
  center}$ =347.680 GHz) and used one more spectral window  
($\nu_{\rm center}$ = 335.570 GHz) to probe the continuum flux
level. 

We started data analysis from calibrated data provided by the Joint ALMA  
Observatory. 
We first checked the visibility plots to locate the targeted emission lines. 
The presence of HCN, HCO$^{+}$, and HNC J=4--3 lines was clearly
recognizable in the visibility plots of individual spectral windows.
In other spectral windows in which we intended to measure the continuum
emission, signatures of emission lines were seen. 
We selected channels that were free of these emission lines to
estimate the continuum flux level.
We then subtracted this continuum level and performed the task
``clean'' for molecular line data.
The ``clean'' procedure was also applied to the continuum data. 
We employed 40 channel spectral binning ($\sim$17 km s$^{-1}$) and 0.1$''$
pixel$^{-1}$ spatial binning in this clean procedure. 

\section{Results}

Figure 1 displays the continuum-a (taken with the HCN and HCO$^{+}$
J=4--3 observations) and continuum-b (taken with HNC J=4--3) maps.
The continuum emission properties are presented in Table 3.
The contribution from thermal free-free emission in star-forming 
HII-regions to the observed
continuum flux ($\sim$10 mJy) at $\sim$345 GHz is estimated to be
only $\sim$3 mJy, from the far-infrared (40--500 $\mu$m) luminosity 
(Table 1), by using the equation (1) of \citet{nak05} and even assuming
that the far-infrared luminosity is dominated by starburst activity,
with no AGN contribution.
Thus, we interpret that the detected continuum emission ($\sim$10 mJy)
at $\sim$345 GHz largely comes from dust emission heated by
dust-obscured energy sources (AGN and/or starburst), with a small
contribution from thermal free-free emission.  

A full spectrum taken with our ALMA observations at the continuum peak
position within the beam size is shown in Figure 2.  
Although we targeted HCN, HCO$^{+}$, and HNC J=4--3 emission
lines, we clearly detected other emission lines that show peaks at the
observed frequencies of $\nu_{\rm obs}$ = 353.88--353.90 GHz and
334.96--335.00 GHz based on Gaussian fits. 
Adopting a redshift of z=0.043, their observed frequencies correspond
to rest-frame frequencies of 
$\nu_{\rm rest}$ = 369.09--369.12 GHz and 349.36--349.41 GHz, which 
we attribute to H$_{2}$S 3(2,1)--3(1,2) ($\nu_{\rm rest}$ = 369.101 GHz) and
CH$_{3}$CN v=0 19(3)--18(3) ($\nu_{\rm rest}$ = 349.393 GHz) 
with a possible contribution from CCH v=0 N=4--3 J=7/2--5/2 ($\nu_{\rm rest}$
= 349.400 GHz), respectively (http://www.splatalogue.net), as seen in
Galactic molecular clouds \citep{sut91,sch97}.   
To our knowledge, ours is the first detection of these lines in
extragalactic sources.  
H$_{2}$S, CH$_{3}$CN and CCH emission lines, albeit at different
rotational (J) transitions, were detected in the very nearby ($<$20 Mpc),
bright starburst galaxies and AGNs \citep{mar06,ala11,ala13}.  
The serendipitous detection of these emission lines in IRAS 20551$-$4250 
at z=0.043 ($\sim$190 Mpc) demonstrates the high sensitivity of ALMA. 

Integrated intensity (moment 0) maps of individual molecular lines are 
created, by combining channels with clearly discernible signals.
These maps are displayed in Figure 3 (left), and their properties are summarized 
in Table 4.
The molecular line peak positions agree with the continuum emission peak
within 1 pixel (0.1'').
Spectra around individual detected molecular lines at the continuum peak
position within the beam size are shown in Figure 3 (right).  
Since continuum emission is well subtracted in the
spectra, the moment 0 maps in Figure 3 (left) 
should reflect the properties of individual molecular gas emission 
lines.
The results of the Gaussian fits for individual molecular lines are 
summarized in Table 4.

Figure 4 shows the intensity-weighted mean velocity (moment 1) and
intensity-weighted velocity dispersion (moment 2) maps for HCN and
HCO$^{+}$ J=4--3 lines.  
A rotating gas motion is seen such that gas in the northeastern
(southwestern) region is redshifted (blueshifted), relative to the
nucleus, suggesting that the molecular gas emission is slightly
spatially extended compared to the synthesized beam.  

\section{Discussion}

\subsection{Vibrationally excited HCN J=4--3 emission line}

In the spectrum of HCO$^{+}$ J=4--3 in Figure 3, a weak emission 
feature is seen at the higher velocity (lower frequency) side of
HCO$^{+}$.  
A magnified view of the spectrum around HCO$^{+}$ J=4--3 is shown in the
right panel of Figure 2. 
The observed frequency of this weak emission peak is 341.51--341.58
GHz (based on Gaussian fit), which corresponds to $\nu_{\rm rest}$ =
356.20--356.27 GHz at z=0.043.  
We attribute this weak line to the vibrationally excited HCN v$_{2}$=1
J=4--3 (l=1f) emission line ($\nu_{\rm rest}$ = 356.256 GHz). 
Since the flux excess, with a peak of $\sim$4.8 mJy, is seen in several
consecutive spectral elements (not one element), it is very unlikely
that the excess feature is caused by a spike noise.
We have carefully investigated the spectrum in Figure 2, and 
found that no spectral elements, other than the identified molecular 
lines, show fluxes exceeding 4 mJy (1$\sigma$ noise level is $\sim$1.2 mJy). 
Since the observed peak frequency of this emission feature 
coincides with the expected frequency of HCN v$_{2}$=1 J=4--3 (l=1f) 
and the observed peak position agrees with the continuum and 
other detected stronger molecular line peak positions,
we judge that the HCN v$_{2}$=1 J=4--3 (l=1f) emission feature is real.

This vibrationally excited HCN v$_{2}$=1 J=4--3 (l=1f) emission line
was recently detected in the external galaxy NGC 4418  
\citep{sak10}.
NGC 4418 is a very nearby luminous infrared galaxy (L$_{\rm IR}$ $\sim$
10$^{11}$L$_{\odot}$) at z=0.007, and a luminous buried
AGN-type energy source with high emission surface brightness was found 
\citep{spo01,eva03,ima04,sak13}.
However, based on only one example, determining whether the vibrational
excitation of HCN in NGC 4418 was an exceptional case or not was impossible.  
From our discovery of this second extragalactic source, we can now show
that HCN vibrational excitation is not exceptionally rare. 
Since the HCN v$_{2}$=1 J=4--3 (l=1f) line is only $\sim$400 km s$^{-1}$
away from the much brighter HCO$^{+}$ J=4--3 emission line at v=0 
(Figure 3, right), 
small HCO$^{+}$ J=4--3 line width is very effective to  
recognize the presence of HCN v$_{2}$=1 J=4--3 (l =1f) emission line, 
by clearly de-blending these emission lines.
In fact, IRAS 20551$-$4250 has the smallest molecular line widths
among six LIRGs observed in our ALMA Cycle 0 program.
We infer that the vibrational excitation of HCN may be common
in AGN-hosting galaxies, even though the detection of the 
HCN v$_{2}$=1 J=4--3 (l=1f) line in galaxies with larger
molecular line widths may be hampered by veiling from the overwhelmingly
brighter HCO$^{+}$ J=4--3 emission line at v=0.   

We estimate the flux of the HCN v$_{2}$=1 J=4--3 (l=1f) emission line to be 
0.39$\pm$0.07 [Jy km s$^{-1}$] based on a Gaussian fit (Table 4).
The HCN J=4--3 v$_{2}$=1 (l=1f) to v=0 flux ratio is $\sim$0.04, which is 
a factor of $\sim$5 smaller than the ratio observed in NGC 4418 \citep{sak10}.

In NGC 4418, the infrared 14 $\mu$m HCN absorption feature was detected
\citep{lah07}.
\citet{sak10} estimated that the energy of the HCN v$_{2}$=1 J=4--3   
(l=1f) emission line was smaller than that of the 14 $\mu$m absorption
line and concluded that the observational results could be explained by
the infrared radiative pumping mechanism. 
Unfortunately, no meaningful constraint on the HCN 14 $\mu$m absorption 
feature is placed for IRAS 20551$-$4250 \citep{lah07}, so that further
discussion is not possible.  
However, as discussed by \citet{sak10}, the energy level of HCN
v$_{2}$=1 (l=1f) level is E/k $\sim$1000 K, which is very difficult to
excite with collisions. 
Infrared radiative pumping by absorbing infrared 14 $\mu$m photons 
is the plausible scenario for the vibrational excitation of HCN.
This mechanism should work effectively in an AGN because infrared
14 $\mu$m continuum is radiated strongly due to AGN-heated hot 
dust emission. 
Thus, detection of the vibrationally excited HCN J=4--3 emission
line is expected more often in an AGN than in a starburst galaxy.
In fact, IRAS 20551$-$4250 shows signatures of a luminous obscured
AGN, including the strong power-law 2--10 keV X-ray emission 
\citep{fra03}, the AGN-type infrared 2.5--8 $\mu$m spectral shape
\citep{ris06,san08,ima10,nar10}, and a higher emission surface
brightness ($>$10$^{13}$ L$_{\odot}$ kpc$^{-2}$) than would be sustained
by star-formation activity \citep{ima11}.  

To proceed with further quantitative discussion, it is desirable to
observe the vibrationally excited v$_{2}$=1 (l=1f) line at the J=3--2
and J=2--1 HCN transitions. 
Also, HCO$^{+}$ and HNC have vibrationally excited 
v$_{2}$=1 J=4--3 (l=1f) lines at $\nu_{\rm rest}$ = 358.242 GHz and
365.147 GHz, respectively, but these lines are outside the frequency
range covered with our ALMA Cycle 0 observations. 
Measurements of the v$_{2}$=1 J=4--3 (l=1f) lines of HCO$^{+}$ and HNC
as well as at J=3--2 and J=2--1 are useful to comprehensively understand 
how infrared radiative pumping excites molecules to the vibrationally
excited level and alters the strengths of the rotational (J) transition
lines at v=0 in individual molecules.      
  
\subsection{Line Flux Ratio}

Figure 5 is the plot of the HCN-to-HCO$^{+}$ J=4--3 and HCN-to-HNC J=4--3
flux ratios in IRAS 20551$-$4250. 
The ratios at four different positions in the starburst galaxy NGC 1614,
taken with our ALMA Cycle 0 observations \citep{ima13}, are overplotted
for comparison.   
IRAS 20551$-$4250 shows a higher HCN-to-HCO$^{+}$ J=4--3 flux ratio than 
the starburst galaxy NGC 1614.
 
Based on observations, HCN-to-HCO$^{+}$ flux ratios have been suggested
to be small ($<$1) in starburst-dominated galaxies, but high ($>$1) in
AGNs at J=1--0 \citep{koh05,ima04,ima06,in06,ima07,kri08,ima09}. 
At J=4--3, the HCN-to-HCO$^{+}$ flux ratio in IRAS 20551$-$4250 is
higher than the starburst galaxy NGC 1614, but the absolute value is 
less than unity. 
Since the critical density of HCN J=4--3 (n$_{\rm crit}$ $\sim$ 
2 $\times$ 10$^{7}$ cm$^{-3}$) is higher than that of HCO$^{+}$ J=4--3 
(n$_{\rm crit}$ $\sim$ 4 $\times$ 10$^{6}$ cm$^{-3}$) \citep{mei07},
HCO$^{+}$ J=4--3 is more easily excited than HCN J=4--3, which could 
reduce the HCN-to-HCO$^{+}$ flux ratio at J=4--3 compared to lower J
transitions. 
HCN-to-HCO$^{+}$ flux ratios higher than unity at J=4--3 or even higher 
transition are found in a high-redshift AGN-dominated source, 
APM 08279$+$5255 \citep{gar06,wei07}, and two nearby LIRGs 
observed with our ALMA Cycle 0 program  
(Imanishi et al. 2013, in preparation).
In IRAS 20551$-$4250, it may be that (1) HCO$^{+}$ is more 
collisionally-excited to J=4--3 than HCN, 
and (2) the effects of infrared radiative pumping in HCN are 
relatively modest, as indicated from the lower HCN v$_{2}$=1 (l=1f) 
to v=0 flux ratio than NGC 4418.
To determine the physical origin of the observed flux ratio and 
excitation condition, we need to obtain data at lower J transition lines. 

\acknowledgments

We thank the anonymous referee for his/her useful comments, and 
E. Mullar, H. Nagai, and K. Saigo for their kind advice
regarding ALMA data analysis.   
We are also grateful to M. Takami, K. Sakamoto, and S. Takakuwa for 
valuable discussions on infrared radiative pumping.   
M.I. is supported by Grants-in-Aid for Scientific Research (no. 22012006).  
This paper makes use of the following ALMA data:
ADS/JAO.ALMA\#2011.0.00020.S . ALMA is a partnership of ESO (representing
its member states), NSF (USA) and NINS (Japan), together with NRC
(Canada) and NSC and ASIAA (Taiwan), in cooperation with the Republic of
Chile. The Joint ALMA Observatory is operated by ESO, AUI/NRAO, and NAOJ.



\clearpage

\begin{deluxetable}{lcrrrrcccl}
\tabletypesize{\scriptsize}
\tablecaption{Observed properties of IRAS 20551$-$4250 
\label{tbl-1}}
\tablewidth{0pt}
\tablehead{
\colhead{Object} & \colhead{Redshift}   & 
\colhead{f$_{\rm 12}$}   & 
\colhead{f$_{\rm 25}$}   & 
\colhead{f$_{\rm 60}$}   & 
\colhead{f$_{\rm 100}$}  & 
\colhead{log L$_{\rm IR}$} & 
\colhead{log L$_{\rm FIR}$} & 
\colhead{Optical}  & \colhead{Energy} \\
\colhead{} & \colhead{}   & \colhead{[Jy]} & \colhead{[Jy]} 
& \colhead{[Jy]} & \colhead{[Jy]}  & \colhead{[L$_{\odot}$]}  &
\colhead{[L$_{\odot}$]} & \colhead{Class} & \colhead{Source}  \\
\colhead{(1)} & \colhead{(2)} & \colhead{(3)} & \colhead{(4)} & 
\colhead{(5)} & \colhead{(6)} & \colhead{(7)} & \colhead{(8)} & 
\colhead{(9)} & \colhead{(10)} 
}
\startdata
IRAS 20551$-$4250 & 0.043 & 0.28 & 1.91 & 12.78 & 9.95  & 12.0 & 11.9 & LI
(HII) & AGN + starburst \\   
\enddata

\tablecomments{
Col.(1): Object name. 
Col.(2): Redshift. 
Col.(3)--(6): f$_{12}$, f$_{25}$, f$_{60}$, and f$_{100}$ are 
{\it IRAS} fluxes at 12 $\mu$m, 25 $\mu$m, 60 $\mu$m, and 100 $\mu$m,
respectively, taken from the IRAS {\it FSC} catalog. 
Col.(7): Decimal logarithm of infrared (8$-$1000 $\mu$m) luminosity
in units of solar luminosity (L$_{\odot}$), calculated with
$L_{\rm IR} = 2.1 \times 10^{39} \times$ D(Mpc)$^{2}$
$\times$ (13.48 $\times$ $f_{12}$ + 5.16 $\times$ $f_{25}$ +
$2.58 \times f_{60} + f_{100}$) [ergs s$^{-1}$] \citep{sam96}.
Col.(8): Decimal logarithm of far-infrared (40$-$500 $\mu$m) luminosity
in units of solar luminosity (L$_{\odot}$), calculated with
$L_{\rm IR} = 2.1 \times 10^{39} \times$ D(Mpc)$^{2}$
$\times$ ($2.58 \times f_{60} + f_{100}$) [ergs s$^{-1}$] \citep{sam96}.
Col.(9): Optical spectral classification. 
This galaxy is classified as a LINER (HII-region); namely, it has no
obvious optical AGN (Seyfert) signature \citep{duc97}.  
Col.(10): Energy source. 
A luminous buried AGN is detected, in addition to starburst activity 
\citep{fra03,ris06,san08,nar10,ima10,ima11}.
}

\end{deluxetable}

\begin{deluxetable}{llcccc}
\tabletypesize{\scriptsize}
\tablecaption{Log of our ALMA Cycle 0 observations \label{tbl-2}}
\tablewidth{0pt}
\tablehead{
\colhead{Line} & \colhead{Date} & \colhead{Antenna} & 
\multicolumn{3}{c}{Calibrator} \\ 
\colhead{} & \colhead{(UT)} & \colhead{Number} &
\colhead{Bandpass} & \colhead{Flux} & \colhead{Phase}  \\
\colhead{(1)} & \colhead{(2)} & \colhead{(3)} & \colhead{(4)} &
\colhead{(5)} & \colhead{(6)}
}
\startdata 
HCN/HCO$^{+}$ J=4--3 & 2012 June 1 & 18 & 3C454.3 & Neptune & J2056$-$472 \\
 & 2012 July 26 & 17 & 3C454.3 & Neptune & J2056$-$472 \\
HNC J=4--3 & 2012 June 2 & 19 & 3C454.3 & Neptune & J2056$-$472 \\
 & 2012 July 26 & 18 & 3C454.3 & Neptune & J2056$-$472 \\
\enddata

\tablecomments{
Col.(1): Observed line.
Col.(2): Observation date (UT). 
Col.(3): Number of antennas used for observations.
Cols.(4), (5), and (6): Bandpass, flux, and phase calibrator for the
target source, respectively.  
}

\end{deluxetable}

\begin{deluxetable}{ccrccl}
\tabletypesize{\scriptsize}
\tablecaption{Continuum emission \label{tbl-3}}
\tablewidth{0pt}
\tablehead{
\colhead{Continuum} & \colhead{Frequency} & \colhead{Flux} & 
\colhead{Peak Coordinate} & \colhead{rms} & \colhead{Synthesized beam} \\
\colhead{} & \colhead{[GHz]} & \colhead{[mJy beam$^{-1}$]} & 
\colhead{(RA,DEC)} & \colhead{[mJy beam$^{-1}$]} & 
\colhead{[arcsec $\times$ arcsec] ($^{\circ}$)} \\  
\colhead{(1)} & \colhead{(2)} & \colhead{(3)}  & \colhead{(4)}  &
\colhead{(5)} & \colhead{(6)}  
}
\startdata 
a & 346.7 & 9.4 (67$\sigma$) & (20 58 26.79, $-$42 39 00.3) &
0.14 & 0.6 $\times$ 0.4 (90$^{\circ}$) \\
b & 341.6 & 10.1 (52$\sigma$) & (20 58 26.80, $-$42 39 00.3) &
0.19 & 0.8 $\times$ 0.4 (91$^{\circ}$) \\  
\enddata

\tablecomments{
Col.(1): Continuum-a or -b. Continuum-a data were taken simultaneously 
with HCN and HCO$^{+}$ J=4--3 observations, and continuum-b data
were obtained at the same time as HNC J=4--3 observations. 
Col.(2): Central frequency in [GHz].
Col.(3): Flux in [mJy beam$^{-1}$] at the emission peak. 
The detection significance relative to the rms noise is shown in parentheses. 
Col.(4): The coordinate of the continuum emission peak in J2000.
Col.(5): The rms noise level (1$\sigma$) in [mJy beam$^{-1}$].
Col.(6): Synthesized beam in [arcsec $\times$ arcsec] and position angle. 
The position angle is 0$^{\circ}$ along the north--south direction
and increases in the counterclockwise direction. 
}

\end{deluxetable}

\clearpage

\begin{deluxetable}{lccl|cccc}
\tabletypesize{\scriptsize}
\tablecaption{Molecular line flux \label{tbl-4}}
\tablewidth{0pt}
\tablehead{
\colhead{Line} & \multicolumn{3}{c}{Integrated intensity (moment 0) map} & 
\multicolumn{4}{c}{Gaussian line fit} \\  
\colhead{} & \colhead{Peak} & \colhead{rms} & \colhead{Beam} &
\colhead{Velocity} & \colhead{Peak} & \colhead{FWHM} & \colhead{Flux} \\ 
\colhead{} & \colhead{[Jy beam$^{-1}$ km s$^{-1}$]} &
\colhead{[Jy beam$^{-1}$ km s$^{-1}$]} & 
\colhead{[$"$ $\times$ $"$] ($^{\circ}$)} &
\colhead{[km s$^{-1}$]} & \colhead{[mJy]} & \colhead{[km s$^{-1}$]} & 
\colhead{[Jy km s$^{-1}$]} \\ 
\colhead{(1)} & \colhead{(2)} & \colhead{(3)} & \colhead{(4)} & 
\colhead{(5)} & \colhead{(6)} & \colhead{(7)} & \colhead{(8)} 
}
\startdata 
HCN  J=4--3& 8.1 (77$\sigma$) & 0.10 & 0.6 $\times$ 0.4
(90$^{\circ}$) & 12894$\pm$1 & 50$\pm$1 & 180$\pm$3 & 9.5$\pm$0.2 \\
HCO$^{+}$ J=4--3 & 12 (85$\sigma$) & 0.14 & 0.6 $\times$ 0.4
(90$^{\circ}$) & 12886$\pm$1 & 70$\pm$1 & 190$\pm$3 & 14$\pm$1 \\  
H$_{2}$S 3(2,1)--3(1,2) & 2.4 (28$\sigma$) & 0.086 & 0.5 $\times$
0.4 (89$^{\circ}$) & 12893$\pm$4 & 18$\pm$1 & 170$\pm$8 & 3.3$\pm$0.2 \\  
HNC J=4--3 & 5.1 (47$\sigma$) & 0.11 & 0.5 $\times$ 0.4
(39$^{\circ}$) & 12890$\pm$2 & 34$\pm$1 & 160$\pm$4 & 5.8$\pm$0.2 \\  
CH$_{3}$CN 19(3)--18(3) \tablenotemark{a} & 4.0 (39$\sigma$)& 0.10 & 0.7 $\times$ 0.4
(92$^{\circ}$) & 12924$\pm$3 & 25$\pm$1 & 170$\pm$6 & 4.4$\pm$0.2 \\  
HCN v$_{2}$=1f J=4--3 & 0.26(5.0$\sigma$) & 0.052 & 0.6 $\times$ 0.4
(90$^{\circ}$) & 12918$\pm$11 & 3.5$\pm$0.6 & 100(best fit; fix) &
0.39$\pm$0.07 \\   
\enddata

\tablenotetext{a}{CCH N=4--3 J=7/2$-$5/2 line may also contribute.} 

\tablecomments{
Col.(1): Observed molecular line. 
HCN, HCO$^{+}$, and H$_{2}$S lines were observed simultaneously with
continuum-a.
HNC and CH$_{3}$CN (+CCH) lines were taken at the same time as
continuum-b. 
Col.(2): Integrated intensity in [Jy beam$^{-1}$ km s$^{-1}$] at the
emission peak. 
The detection significance relative to the rms noise of the moment 0
maps is given in parentheses. 
Col.(3): The rms noise level (1$\sigma$) in the moment 0 map in 
[Jy beam$^{-1}$ km s$^{-1}$]. 
It varies among different lines, depending on the number of channels 
combined.
Col.(4): Synthesized beam in [arcsec $\times$ arcsec] and position angle. 
The position angle is 0$^{\circ}$ along the north--south direction
and increases in the counterclockwise direction. 
Cols.(5)--(8): Gaussian fits of detected emission lines in the spectra at the
continuum peak position, within the beam size.
Col.(5): Central velocity in [km s$^{-1}$].
Col.(6): Peak flux in [mJy].
Col.(7): Full width at half maximum (FWHM) in [km s$^{-1}$].
Col.(8): Flux in [Jy km s$^{-1}$].
}

\end{deluxetable}

\clearpage

\begin{figure}
\begin{center}
\includegraphics[angle=0,scale=.5]{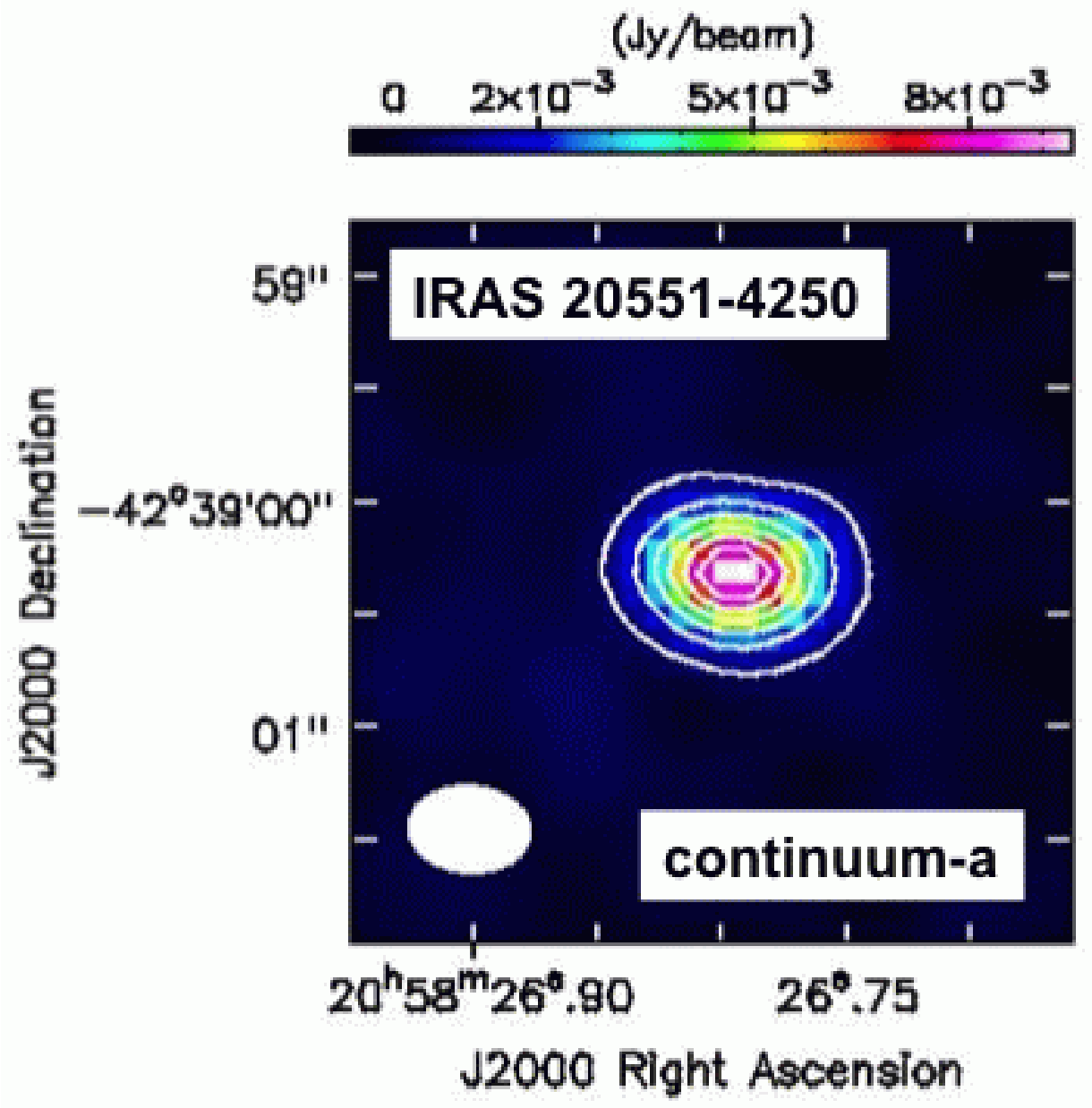} 
\includegraphics[angle=0,scale=.5]{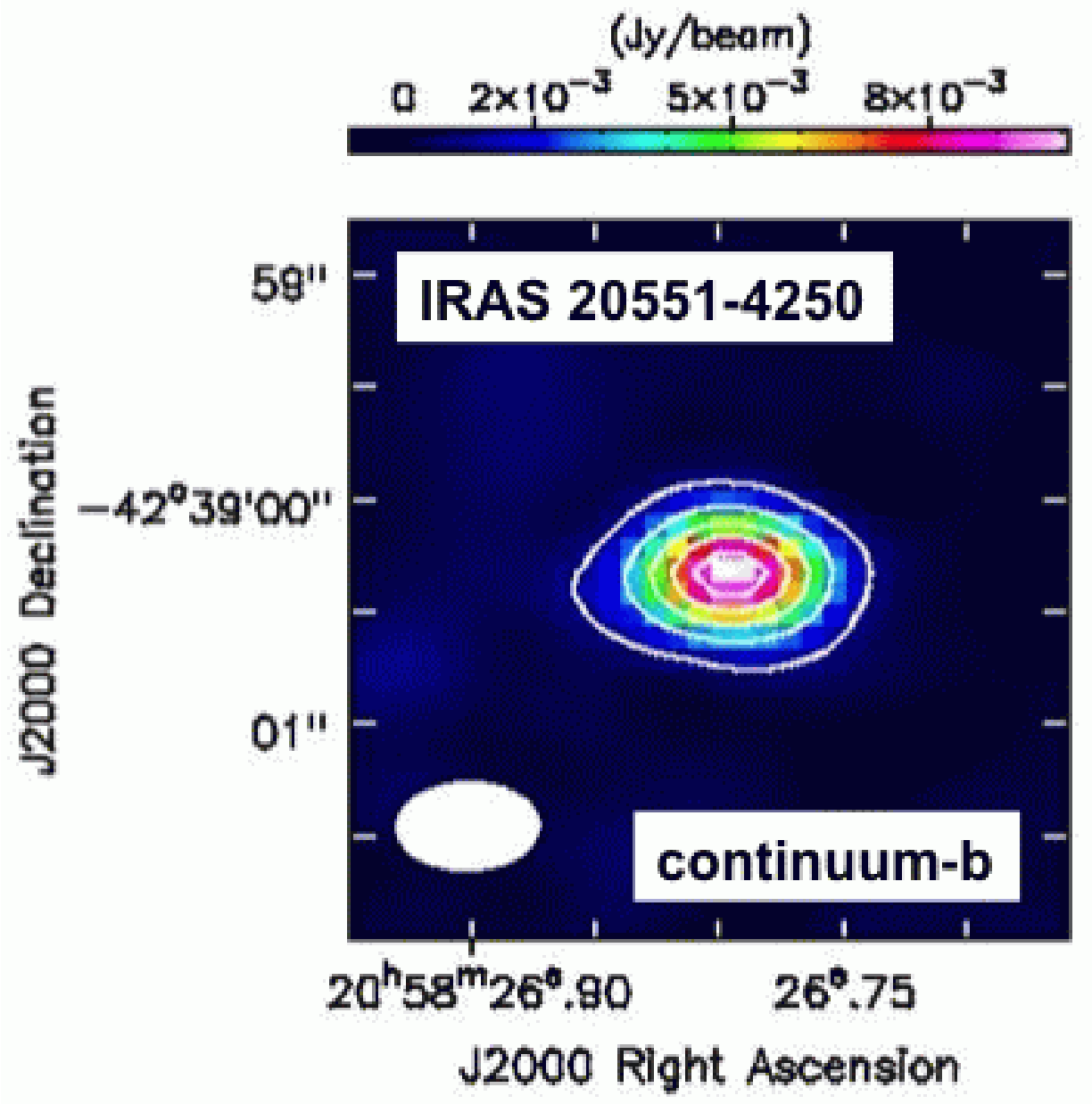} \\
\end{center}
\caption{
Continuum-a ($\sim$346.7 GHz) and -b ($\sim$341.6 GHz) data taken
during the observations of HCN/HCO$^{+}$ J=4--3 and HNC J=4--3,
respectively.  
The contours represent the 5$\sigma$, 15$\sigma$, 25$\sigma$,
35$\sigma$, 45$\sigma$, and 55$\sigma$ levels for continuum-a and the  
5$\sigma$, 15$\sigma$, 25$\sigma$, 35$\sigma$, and 45$\sigma$ levels for
continuum-b.  
}
\end{figure}

\begin{figure}
\begin{center}
\includegraphics[angle=0,scale=.4]{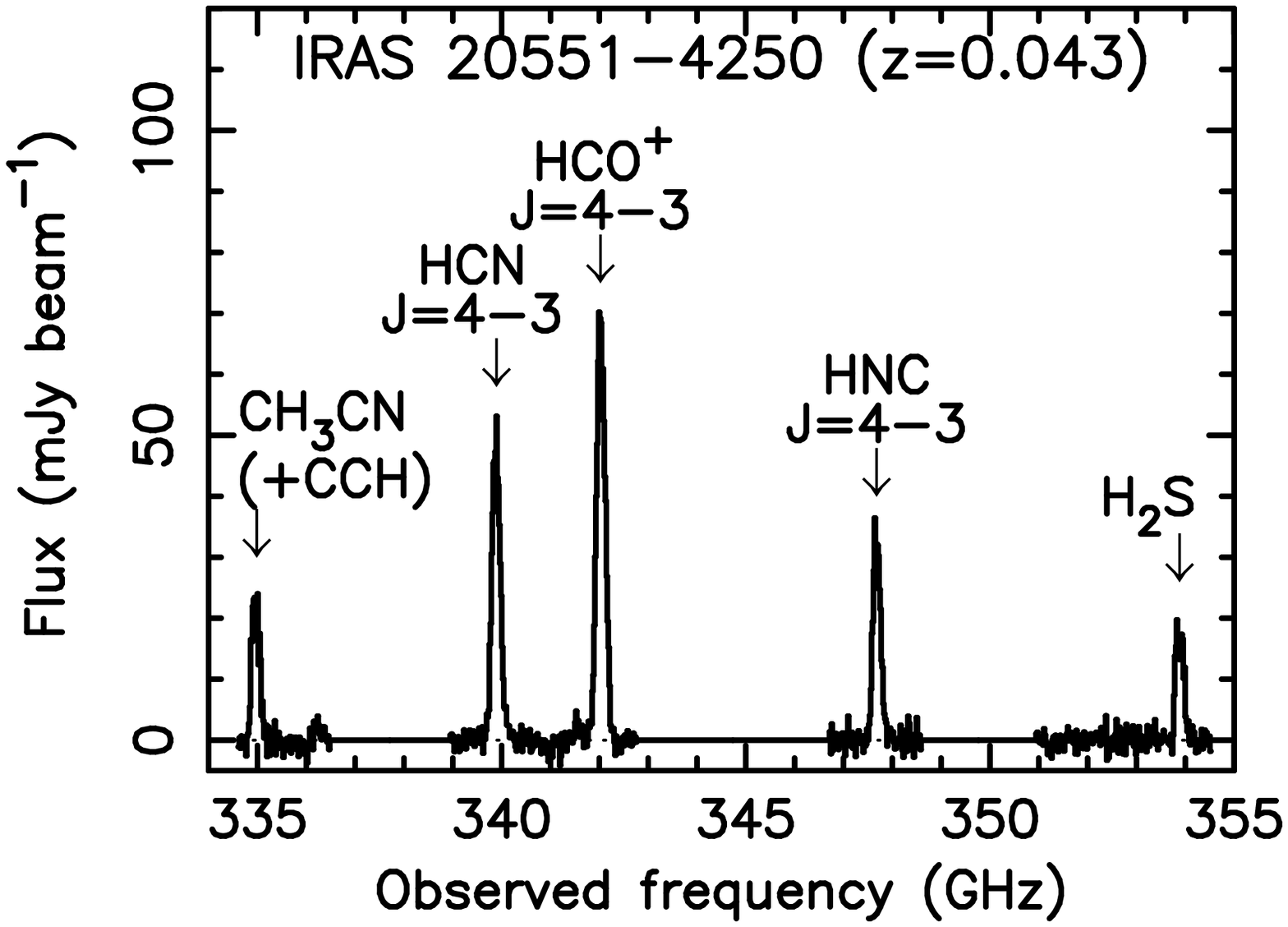} 
\includegraphics[angle=0,scale=.4]{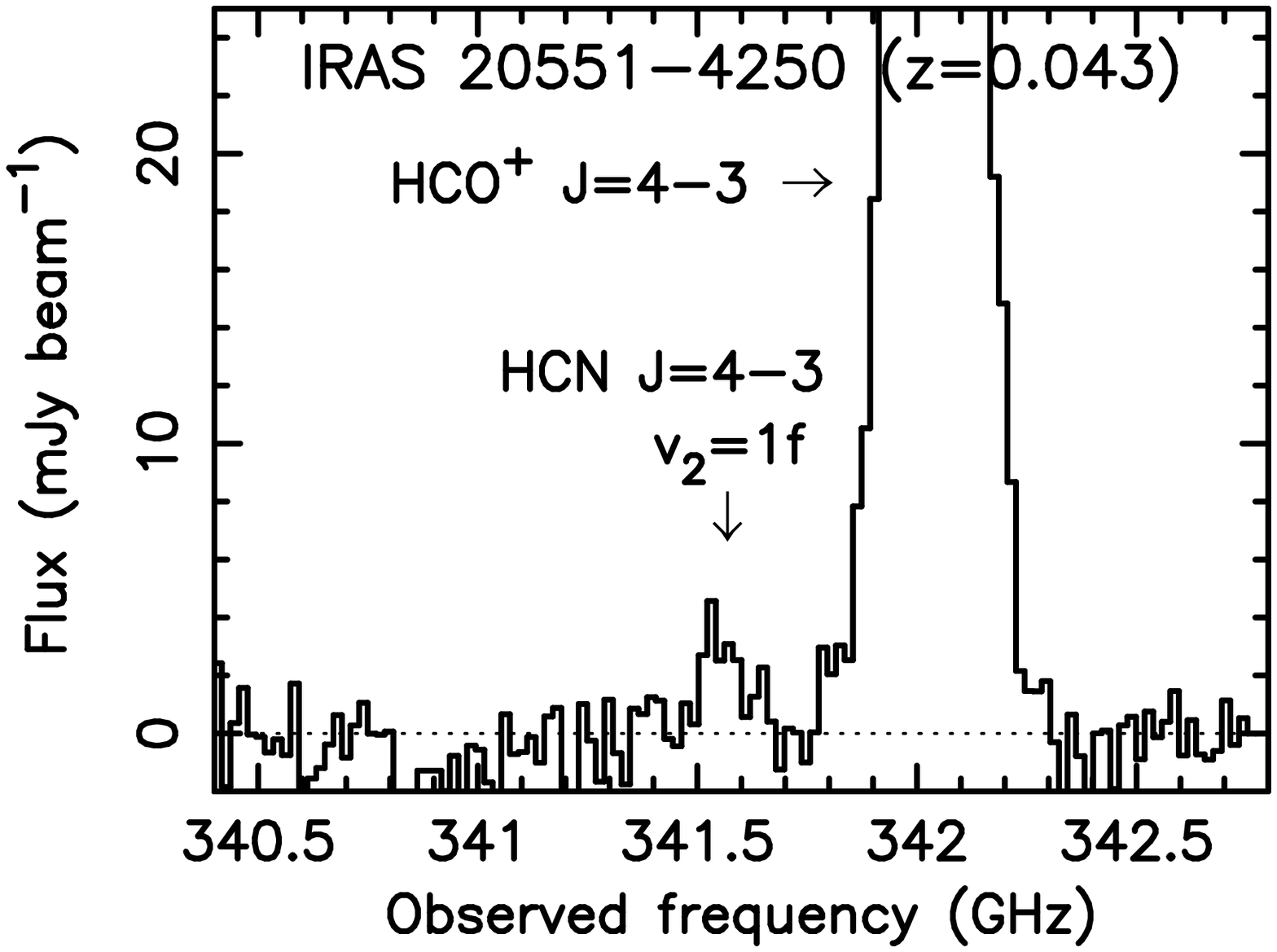}  \\
\end{center}
\caption{
(Left): Full spectrum at the continuum peak position within the beam 
size.   
(Right): Magnified spectrum around the HCO$^{+}$ J=4--3 emission line.
The abscissa is the observed frequency in [GHz] and the ordinate is flux
in [mJy beam$^{-1}$].
}
\end{figure}

\begin{figure}
\begin{center}
\includegraphics[angle=0,scale=.45]{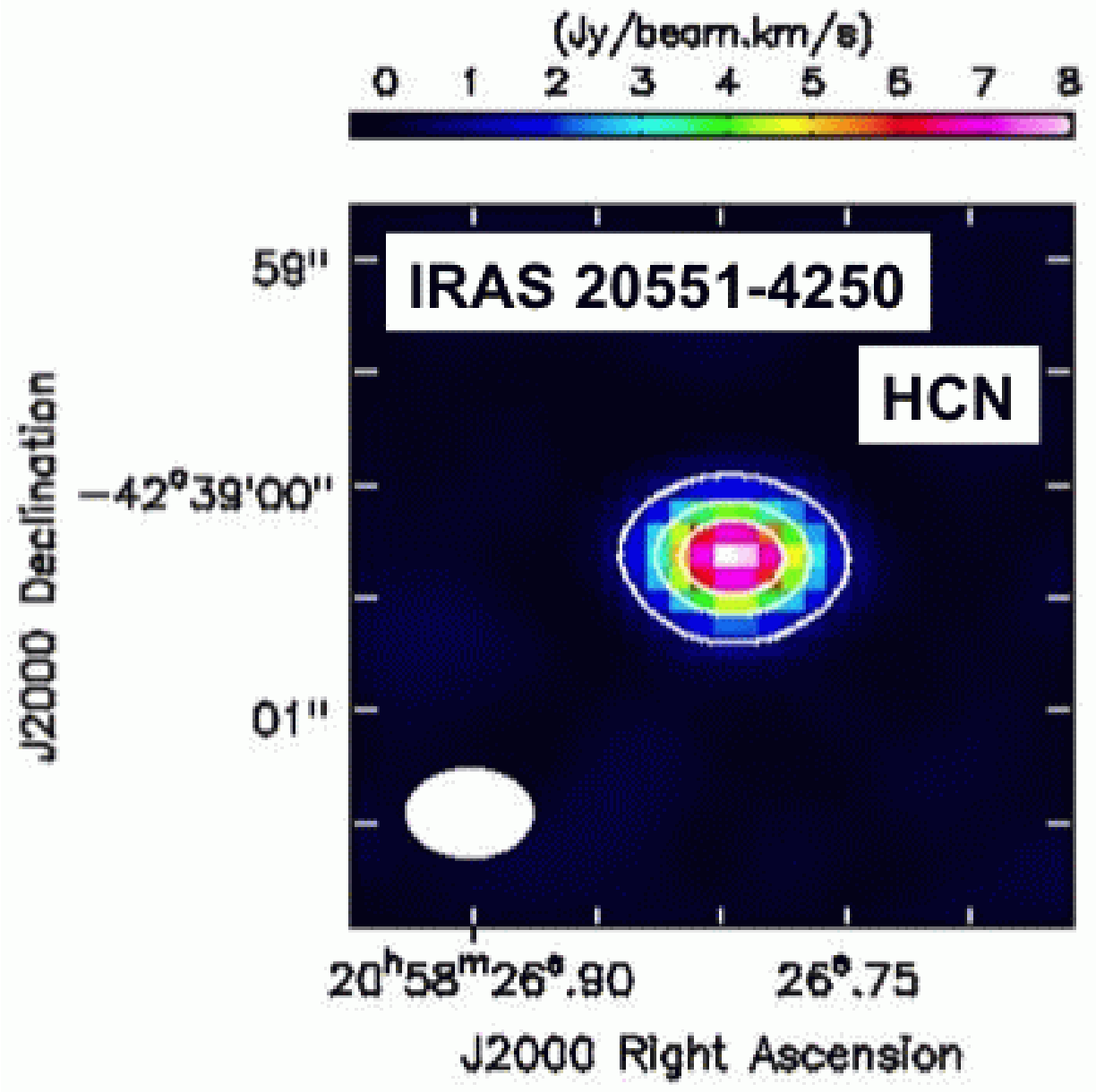} 
\includegraphics[angle=0,scale=.4]{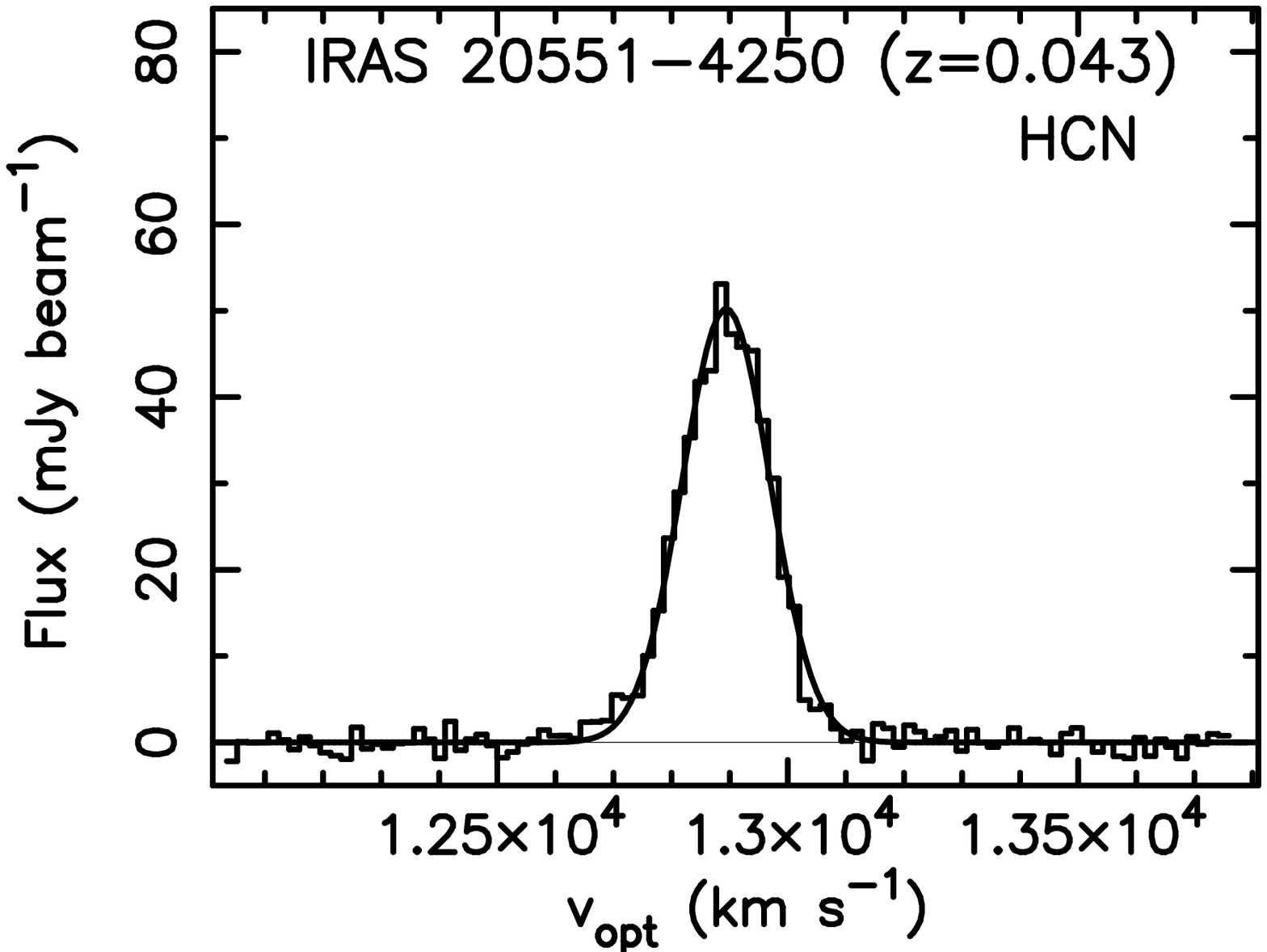}\\ 
\vspace{-1cm}
\includegraphics[angle=0,scale=.45]{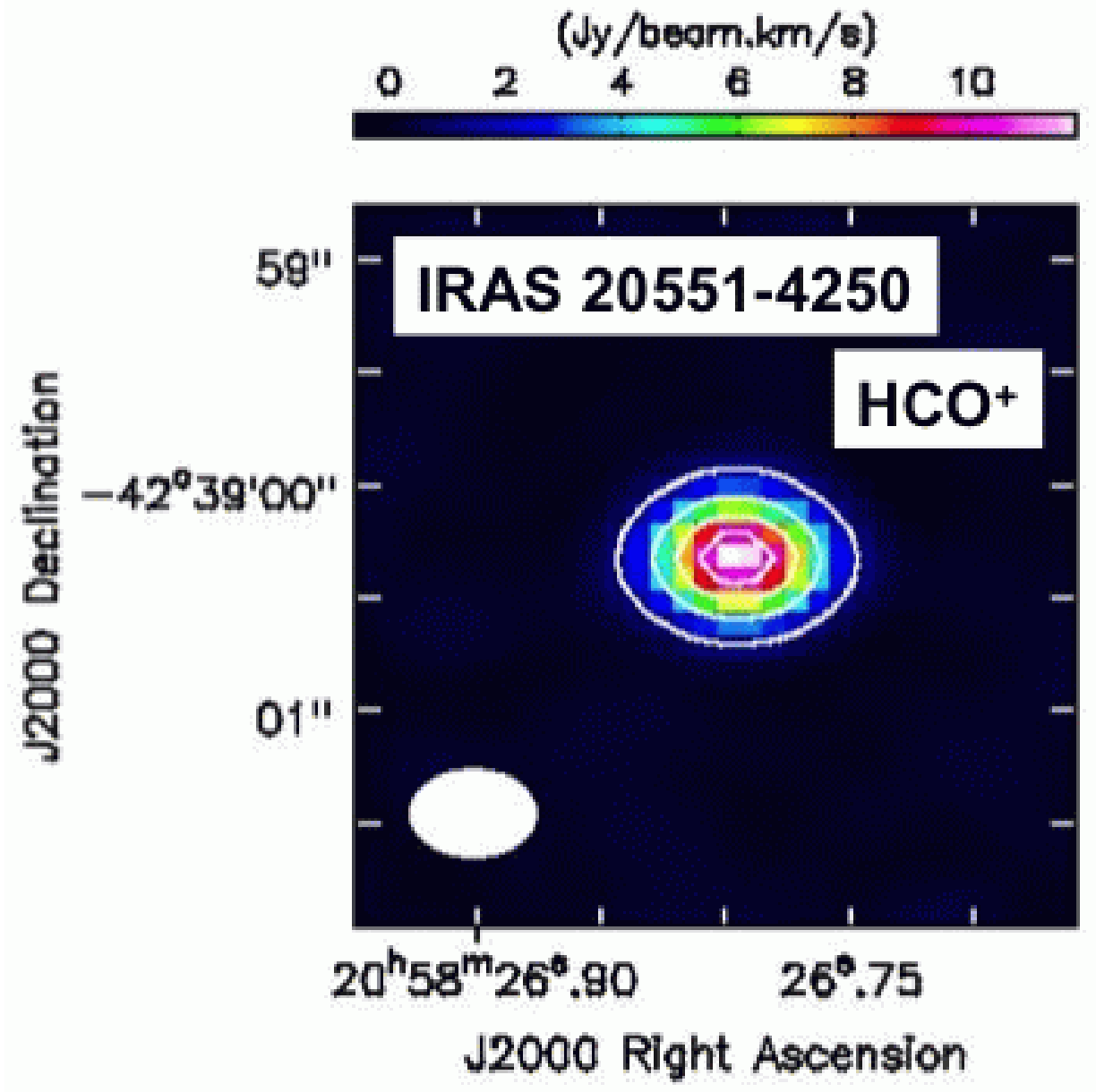} 
\includegraphics[angle=0,scale=.4]{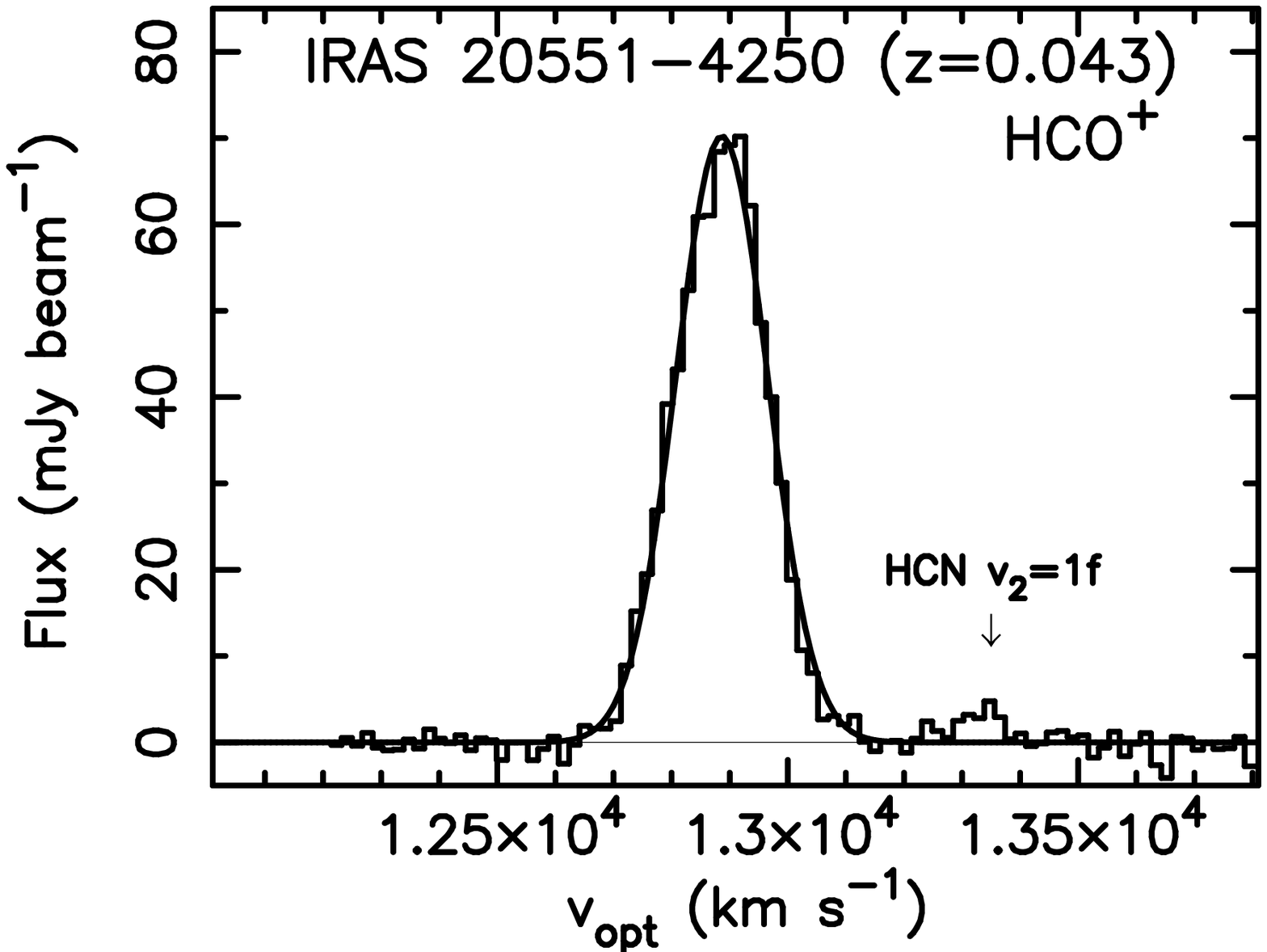}\\ 
\vspace{-1cm}
\includegraphics[angle=0,scale=.45]{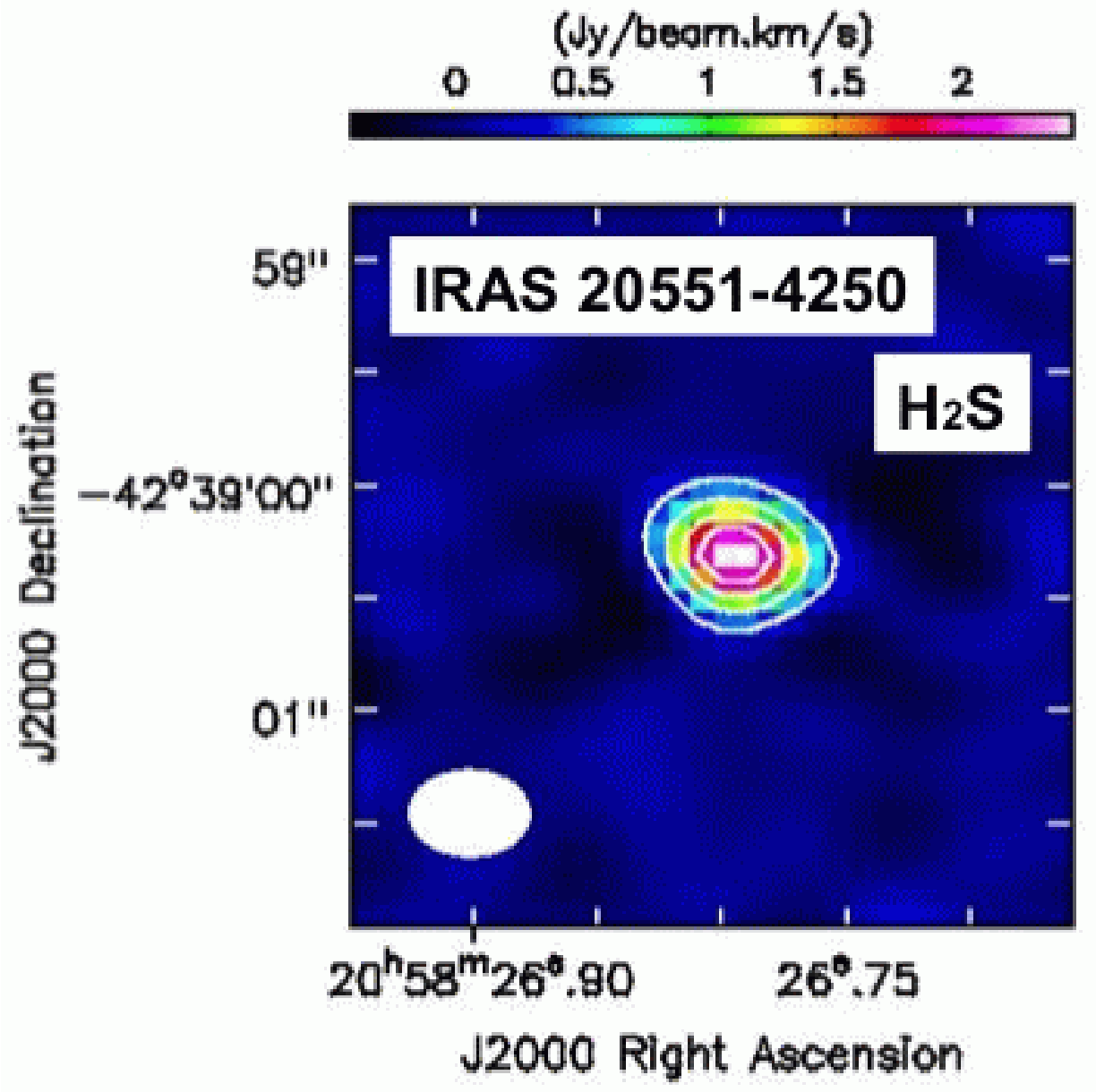} 
\includegraphics[angle=0,scale=.4]{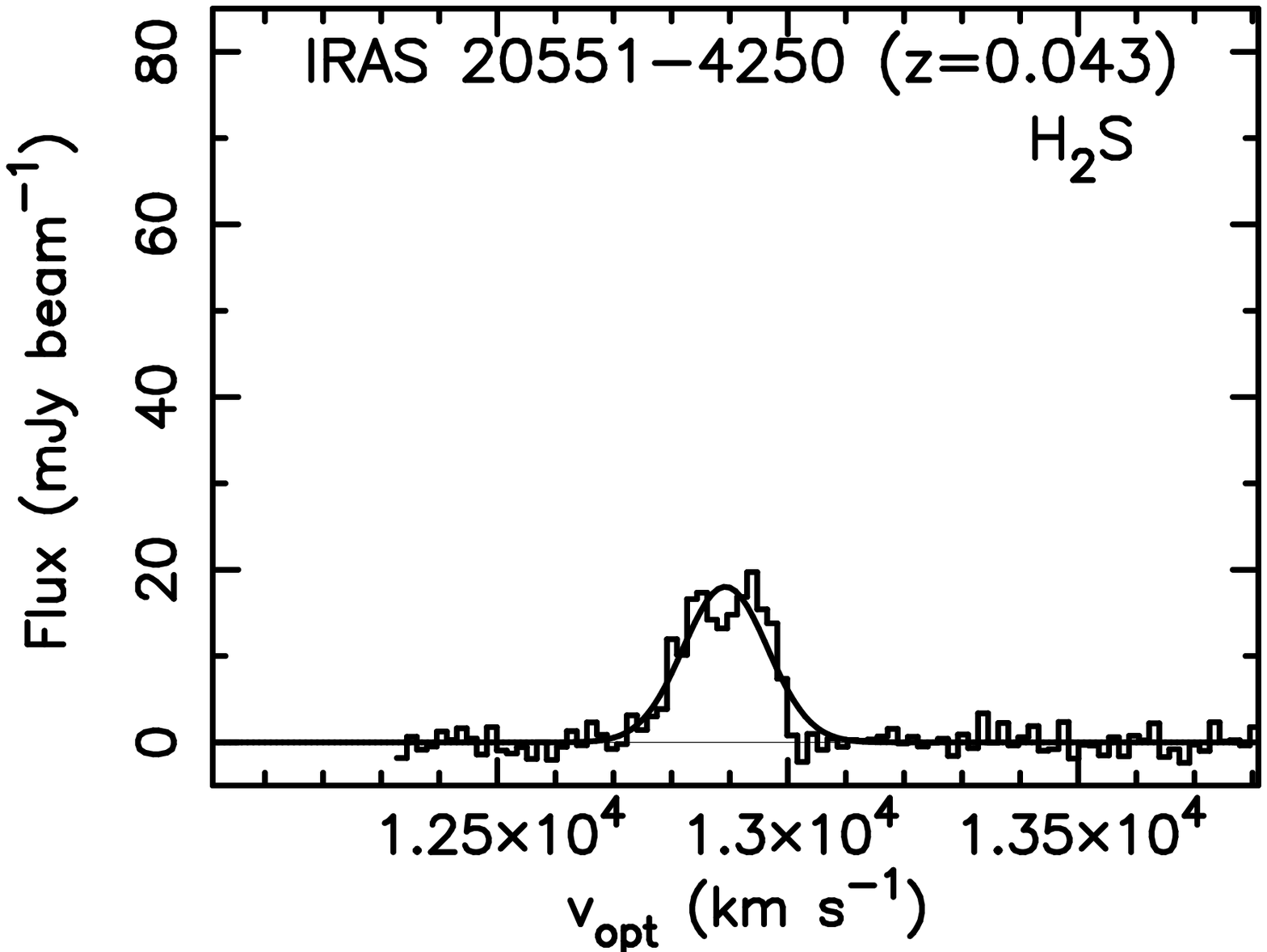}\\
\end{center}
\end{figure}
 
\begin{figure}
\begin{center}
\vspace{-2cm}
\includegraphics[angle=0,scale=.45]{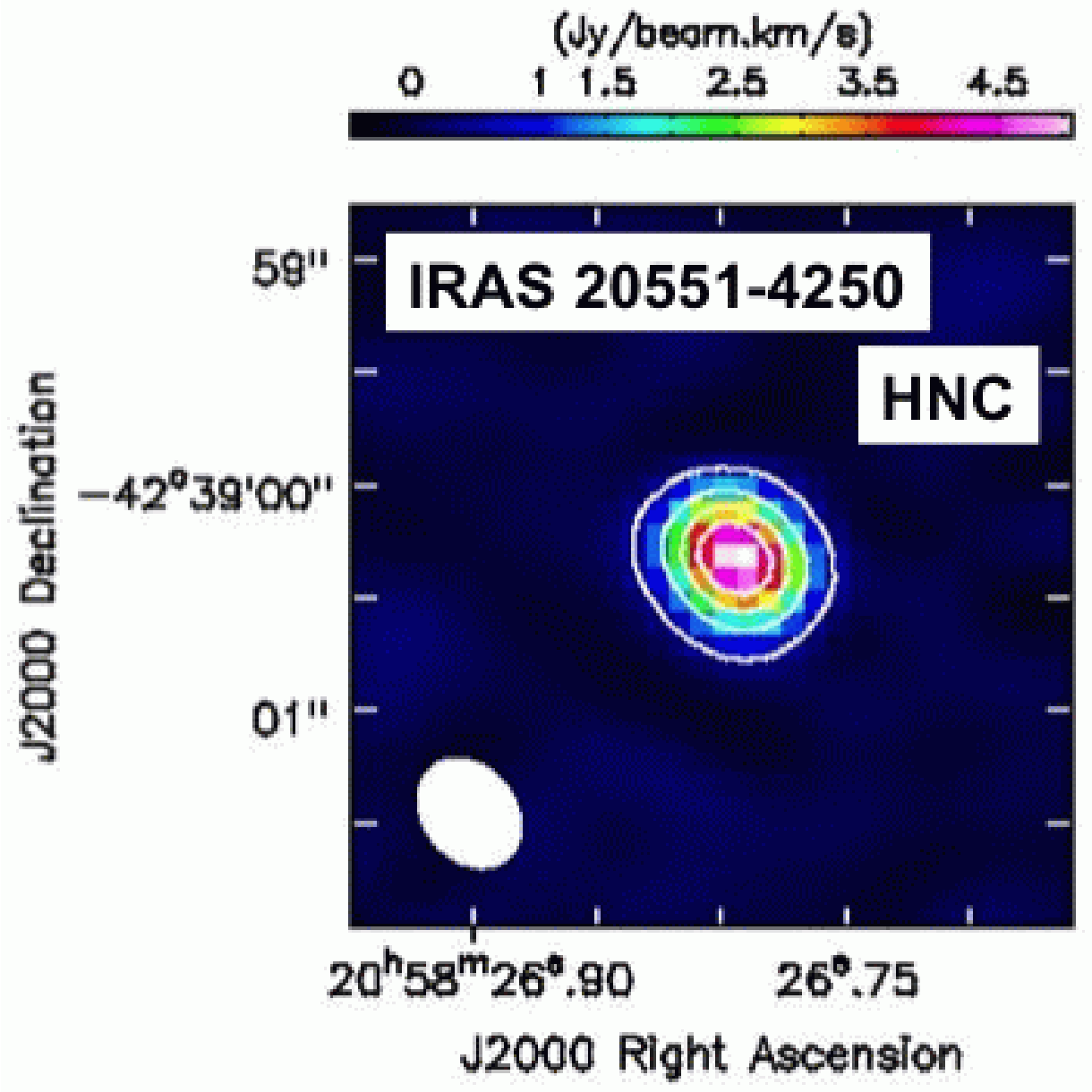} 
\includegraphics[angle=0,scale=.4]{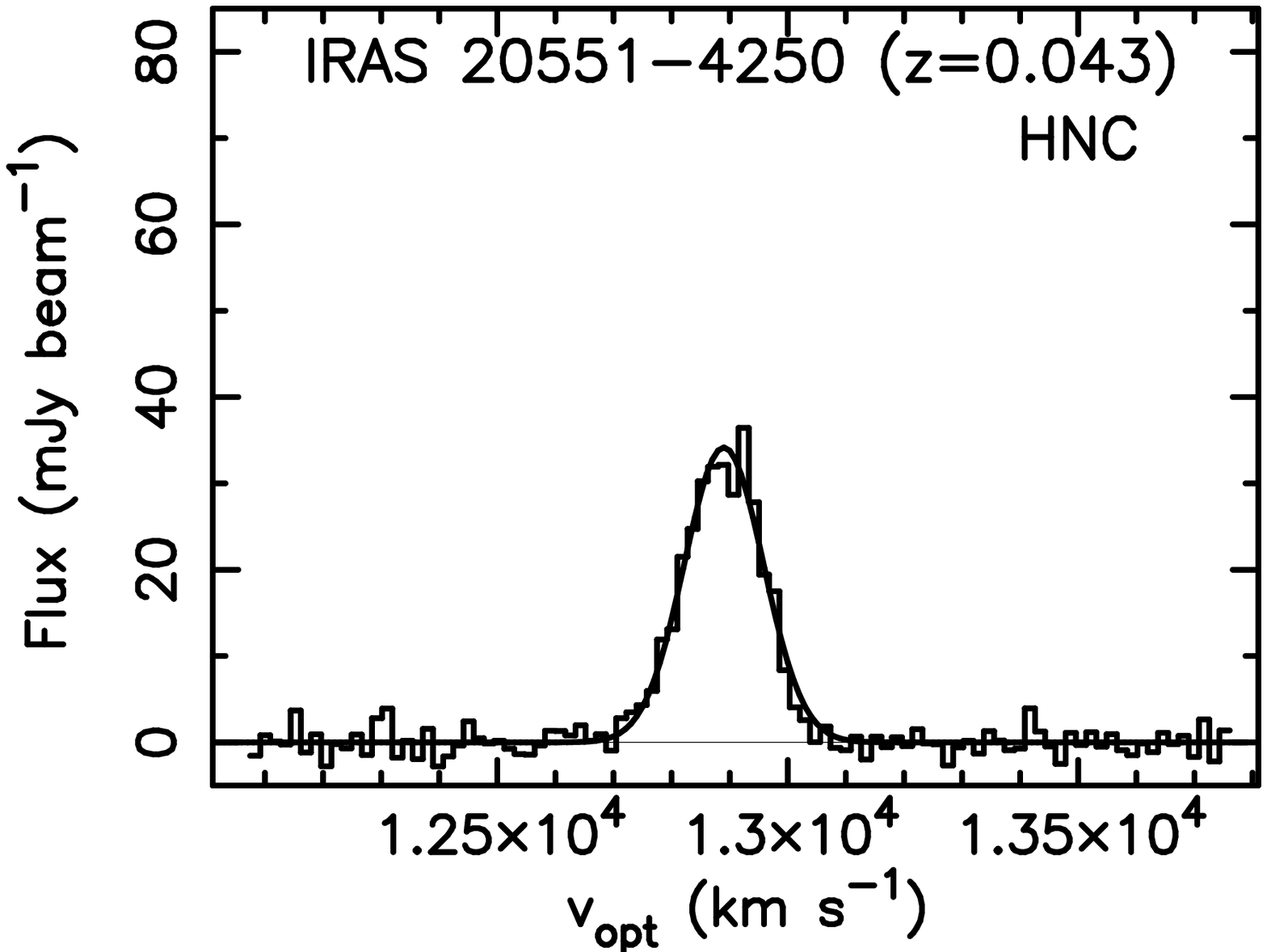}\\ 
\vspace{-1.35cm}
\includegraphics[angle=0,scale=.45]{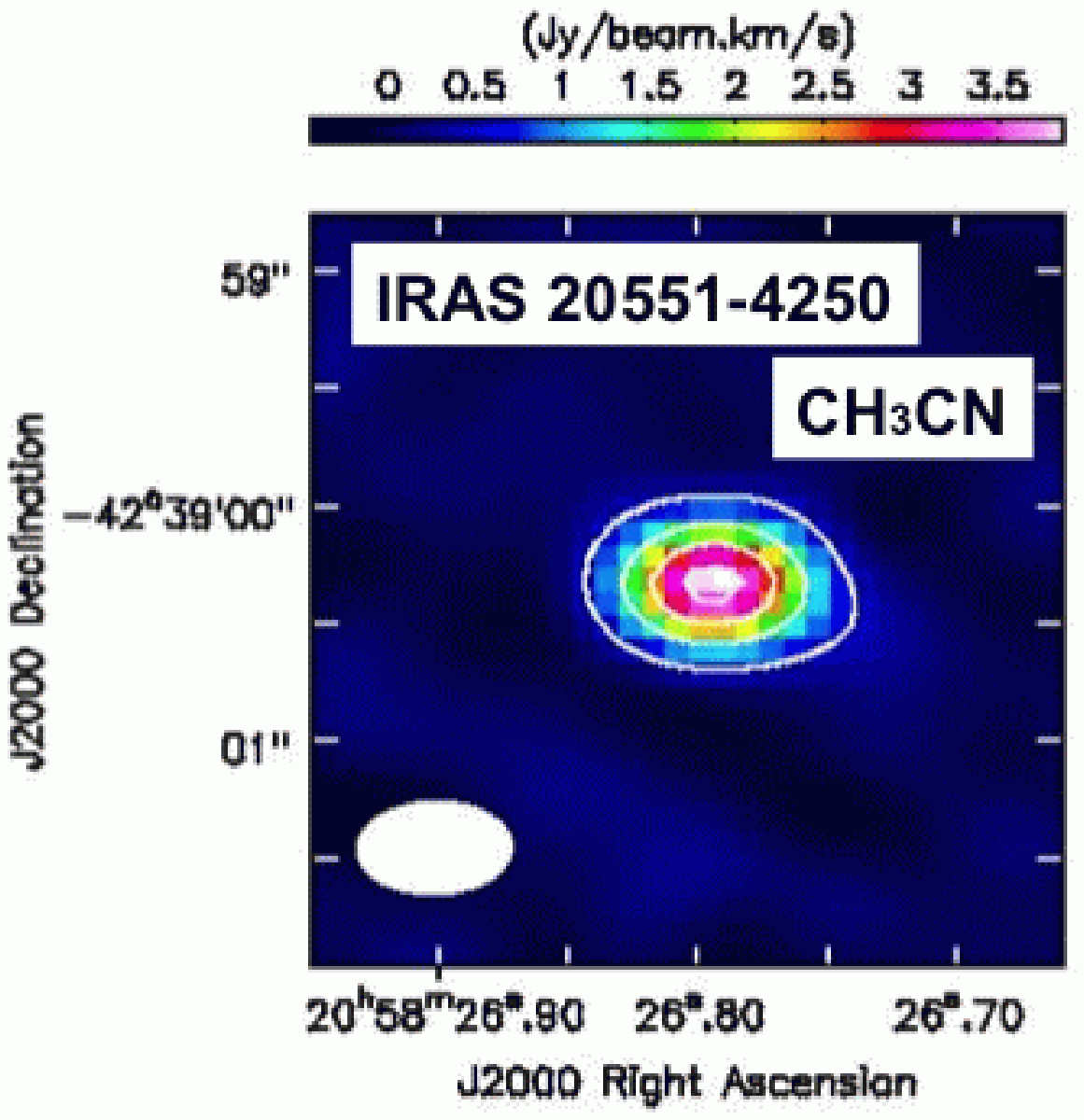} 
\includegraphics[angle=0,scale=.4]{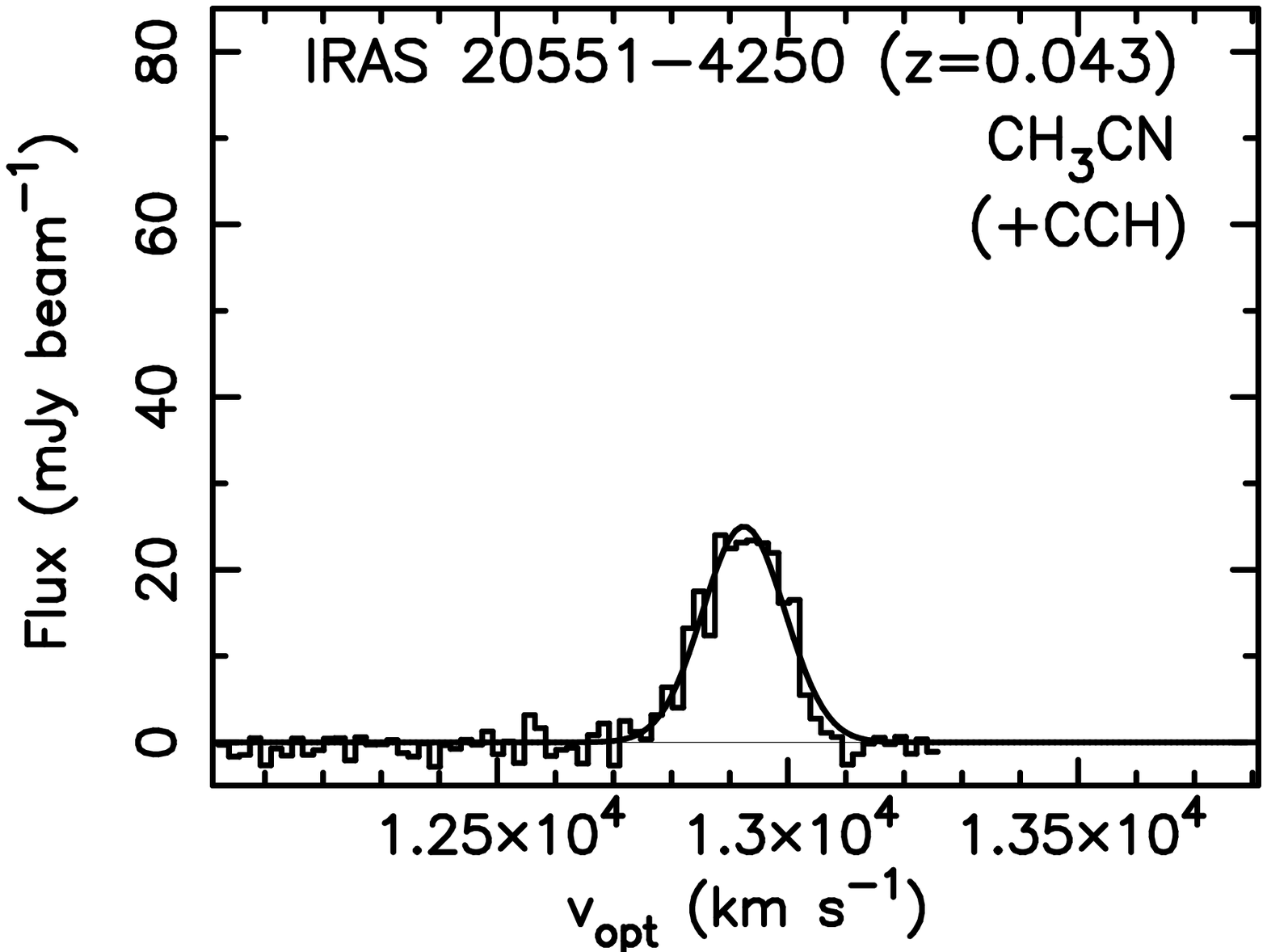}\\ 
\vspace{-1.35cm}
\includegraphics[angle=0,scale=.45]{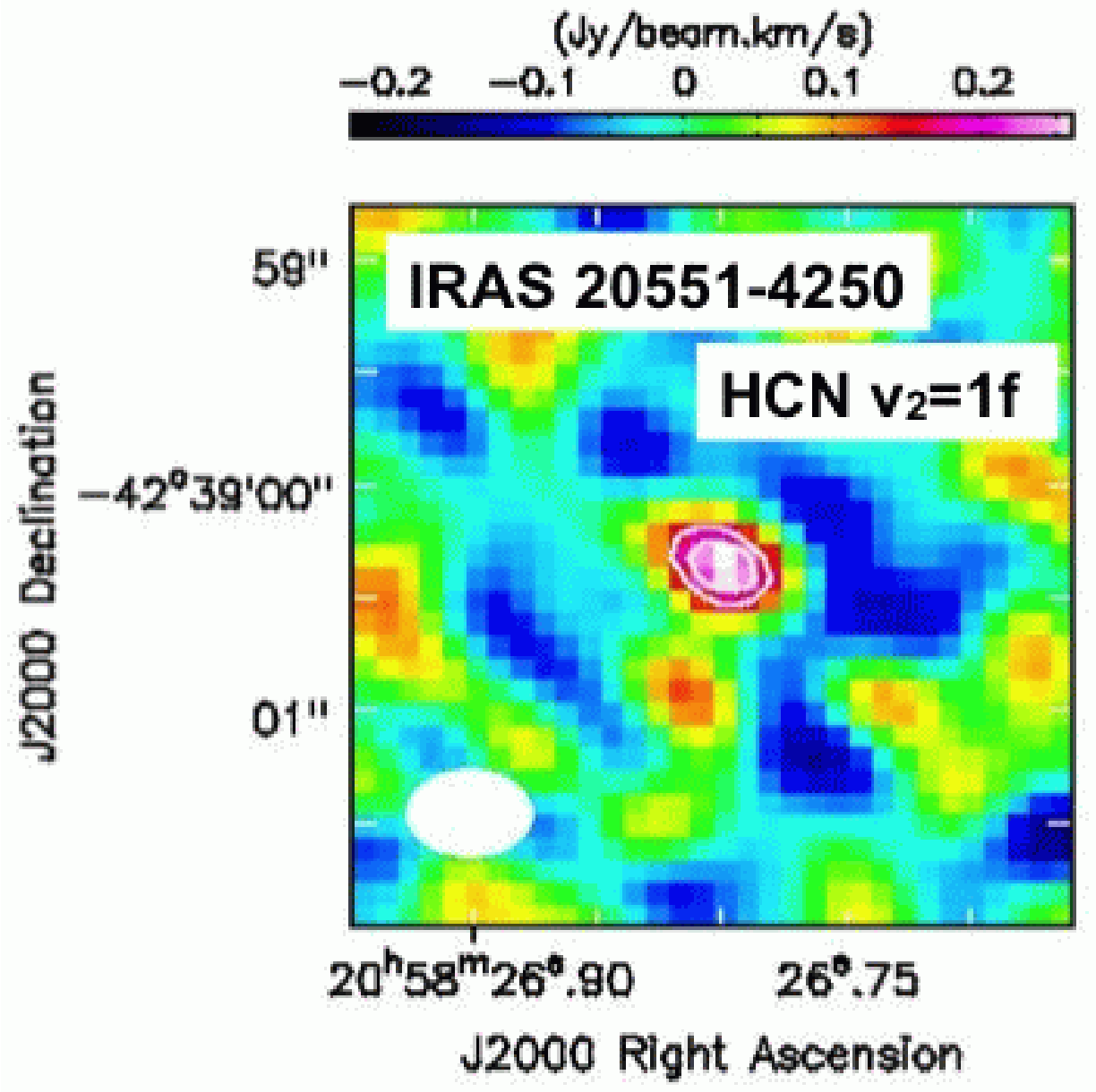} 
\includegraphics[angle=0,scale=.4]{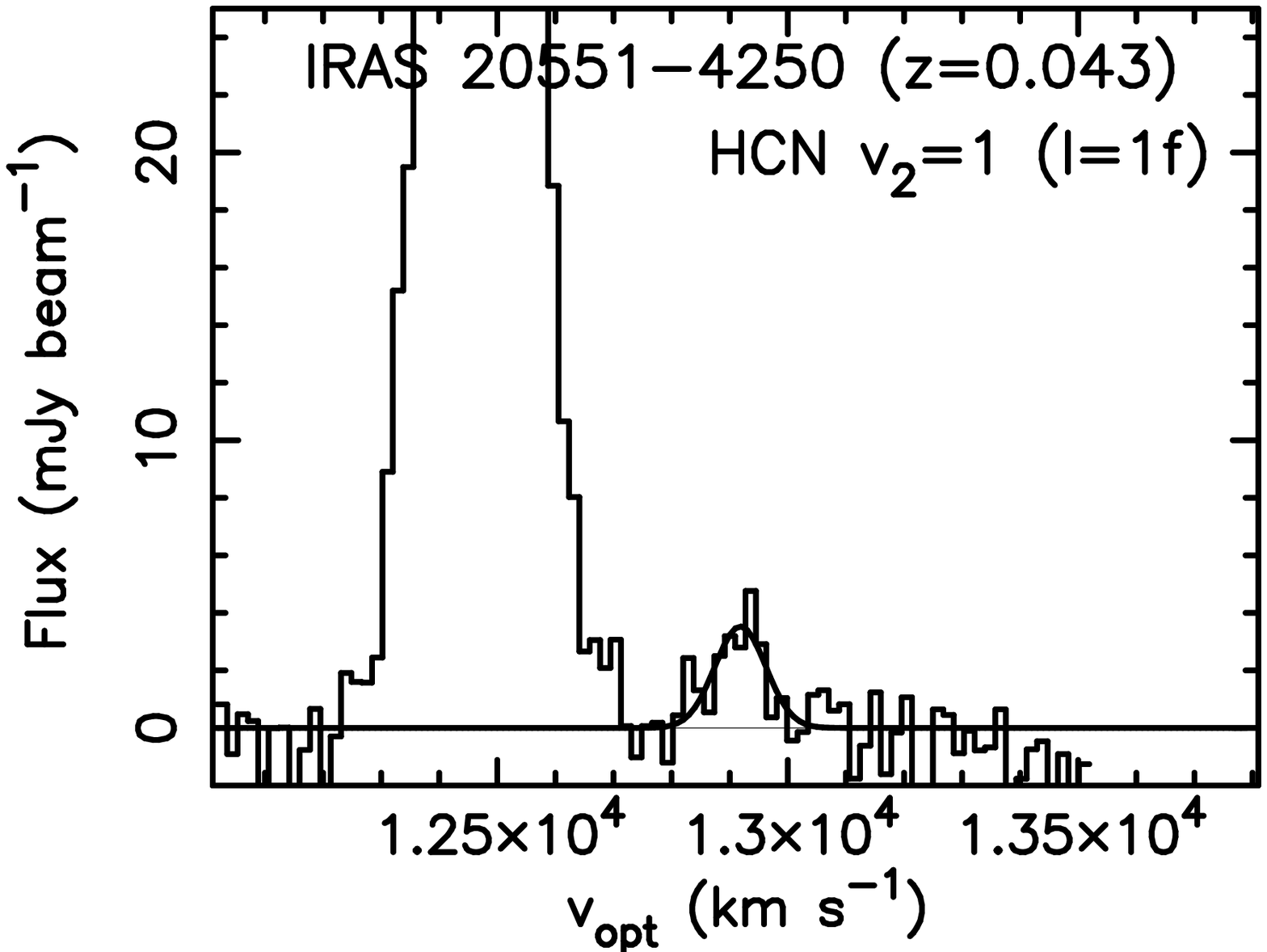}\\ 
\end{center}
\caption{
Integrated intensity (moment 0) maps (left) and spectra at the continuum
peak position, within the beam size (right), of the detected molecular lines 
in IRAS 20551$-$4250. 
HCN, HCO$^{+}$, and H$_{2}$S spectra are extracted 
at the continuum-a peak position.
HNC and CH$_{3}$CN (+CCH) spectra are created at the continuum-b peak.
The contours represent 10$\sigma$, 30$\sigma$, 50$\sigma$ for HCN, 
10$\sigma$, 30$\sigma$, 50$\sigma$, 70$\sigma$ for HCO$^{+}$, 
4$\sigma$, 10$\sigma$, 16$\sigma$, 22$\sigma$ for H$_{2}$S, 
5$\sigma$, 15$\sigma$, 25$\sigma$, 35$\sigma$ for HNC, 
5$\sigma$, 15$\sigma$, 25$\sigma$, 35$\sigma$ for CH$_{3}$CN, and 
3$\sigma$, 4$\sigma$ for HCN v$_{2}$=1 (l=1f).
The 1$\sigma$ levels are different for different lines.
They are summarized in Table 4.
For the spectra, the abscissa is 
optical LSR velocity (v$_{\rm opt}$ $\equiv$ c 
($\lambda$-$\lambda_{\rm 0}$)/$\lambda_{\rm 0}$), and the ordinate is  
flux in [mJy beam$^{-1}$].  
}
\end{figure}

\begin{figure}
\begin{center}
\includegraphics[angle=0,scale=.5]{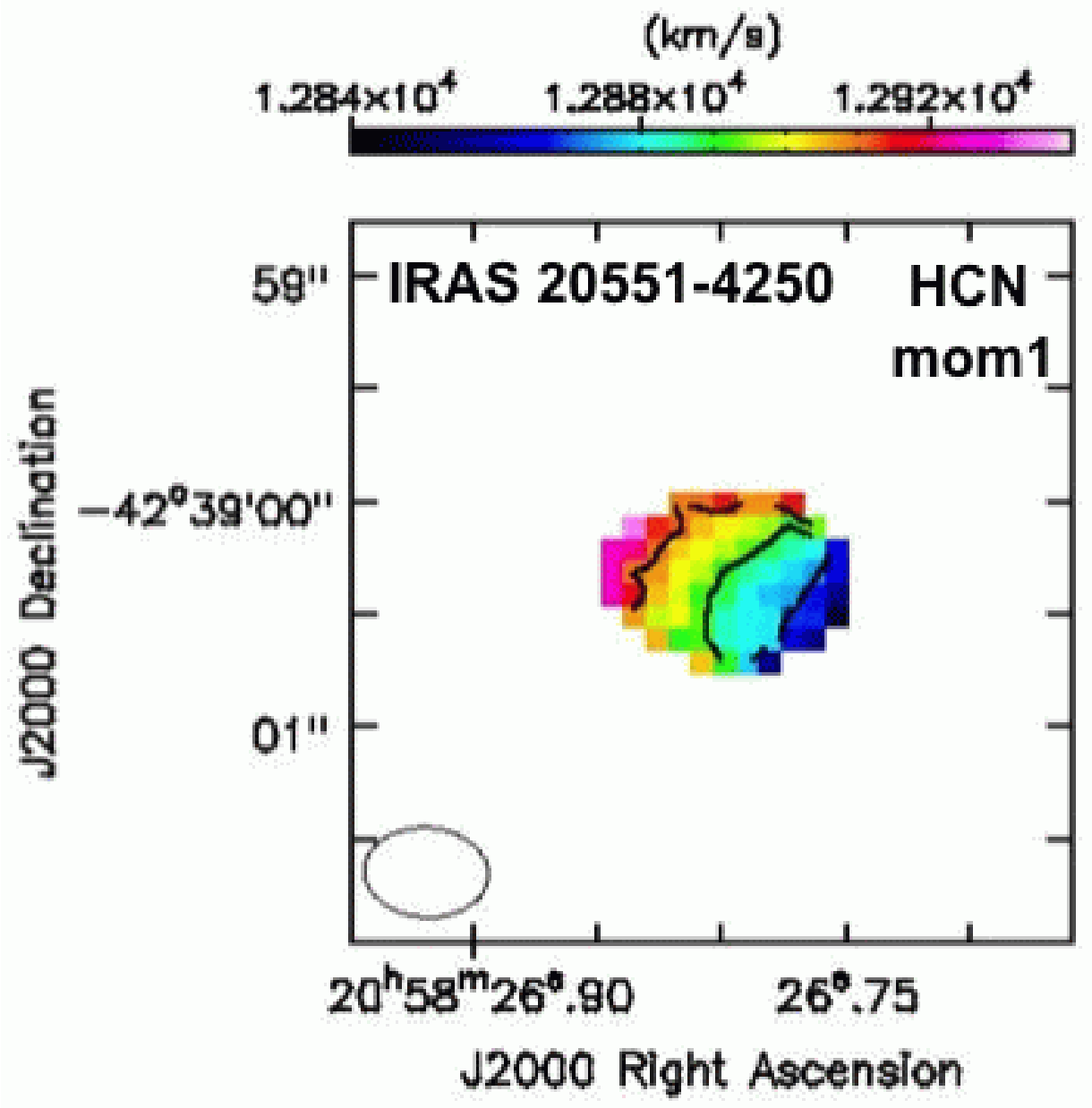} 
\includegraphics[angle=0,scale=.5]{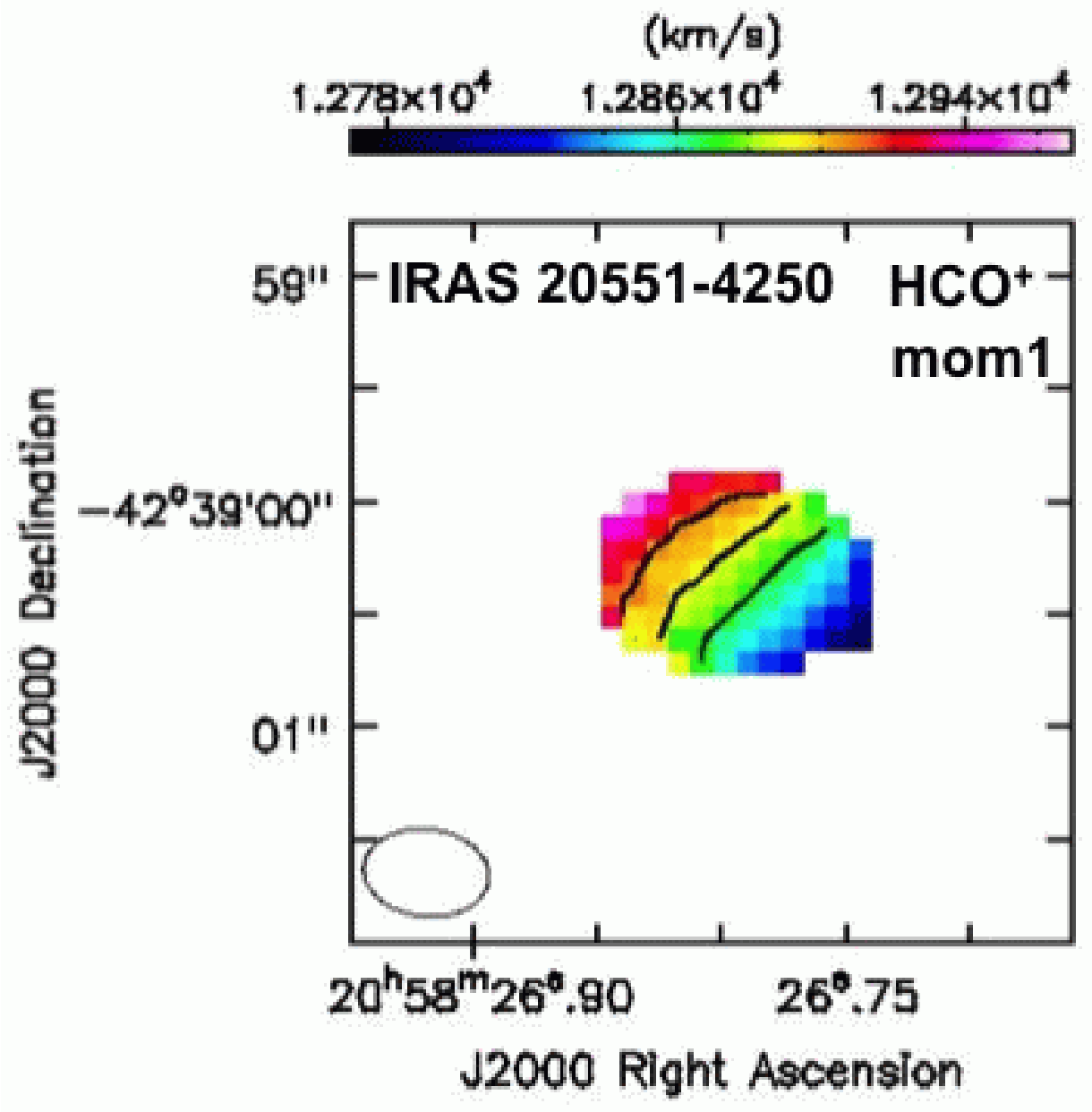} \\
\includegraphics[angle=0,scale=.5]{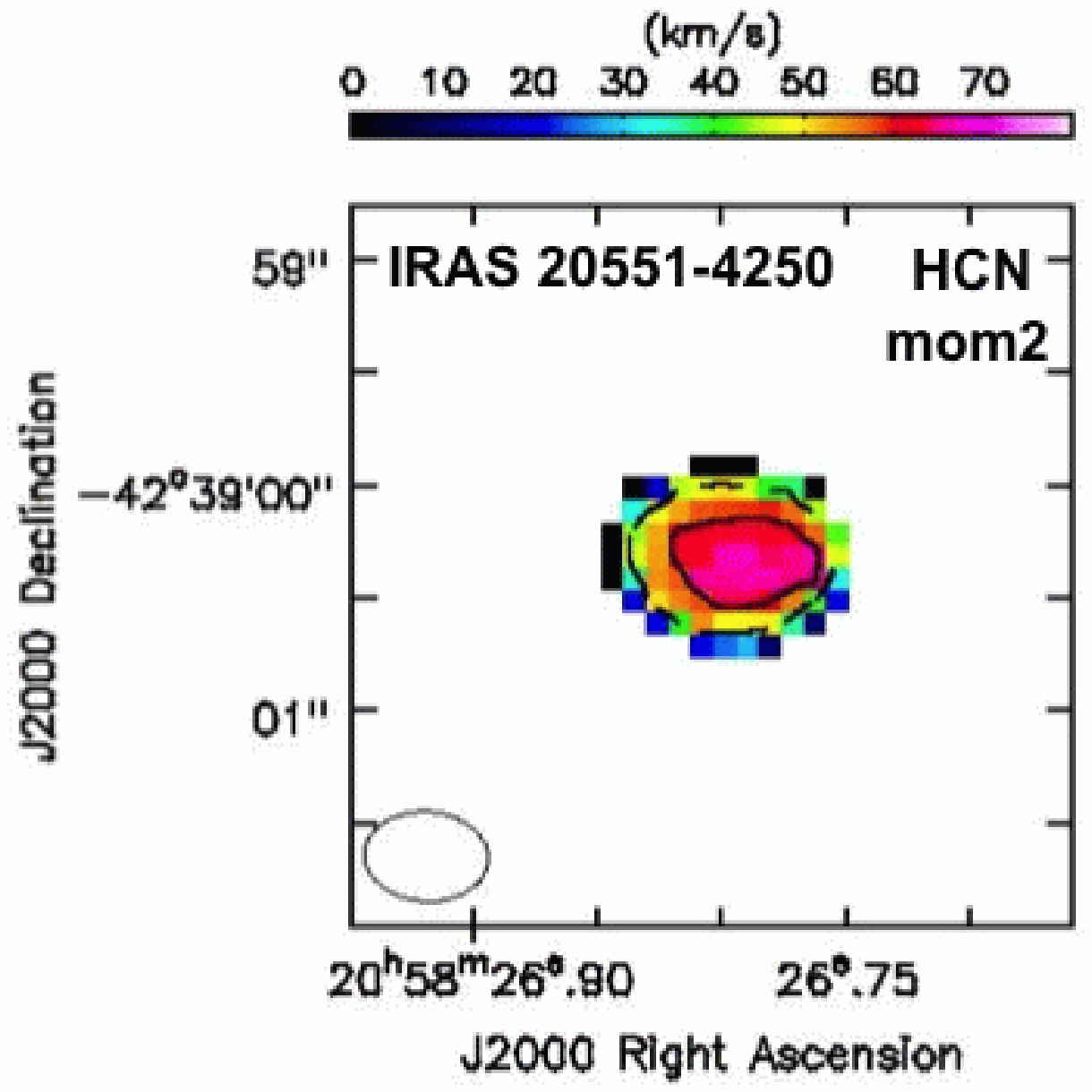} 
\includegraphics[angle=0,scale=.5]{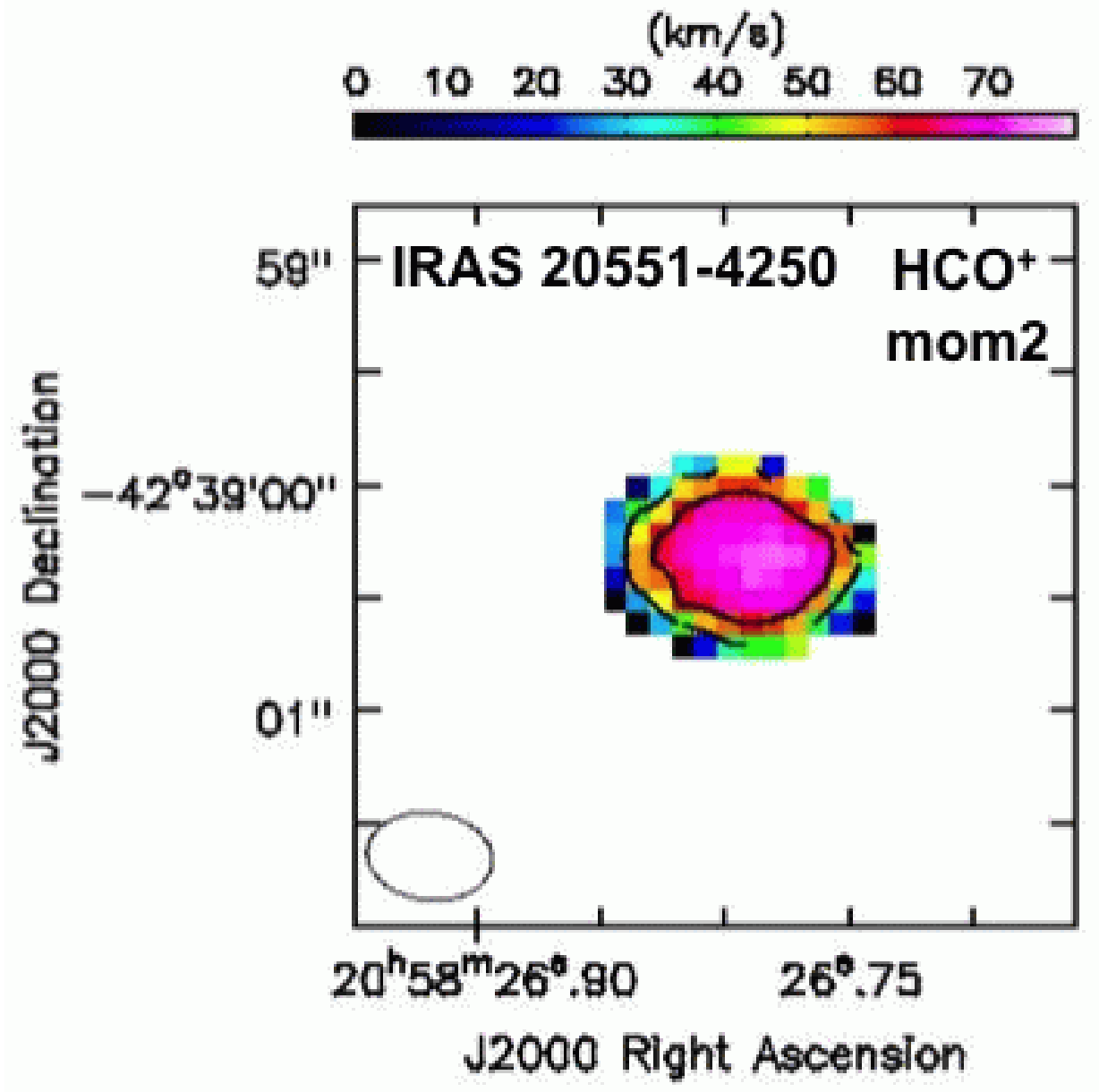} 
\end{center}
\caption{
Intensity-weighted mean velocity (moment 1) and intensity-weighted
velocity dispersion (moment 2) maps of HCN J=4--3 and HCO$^{+}$ J=4--3
emission lines for IRAS 20551$-$4250. 
For moment 1 maps, the velocity is in optical LSR velocity 
(v$_{\rm opt}$ $\equiv$ c ($\lambda$-$\lambda_{\rm 0}$)/$\lambda_{\rm 0}$). 
The contours in moment 1 maps are 12870, 12890, and 12910 km s$^{-1}$
for both HCN and HCO$^{+}$ J=4--3.
The contours in moment 2 maps are 40 and 60 km s$^{-1}$ for both HCN and
HCO$^{+}$ J=4--3. 
}
\end{figure}

\begin{figure}
\begin{center}
\includegraphics[angle=0,scale=.5]{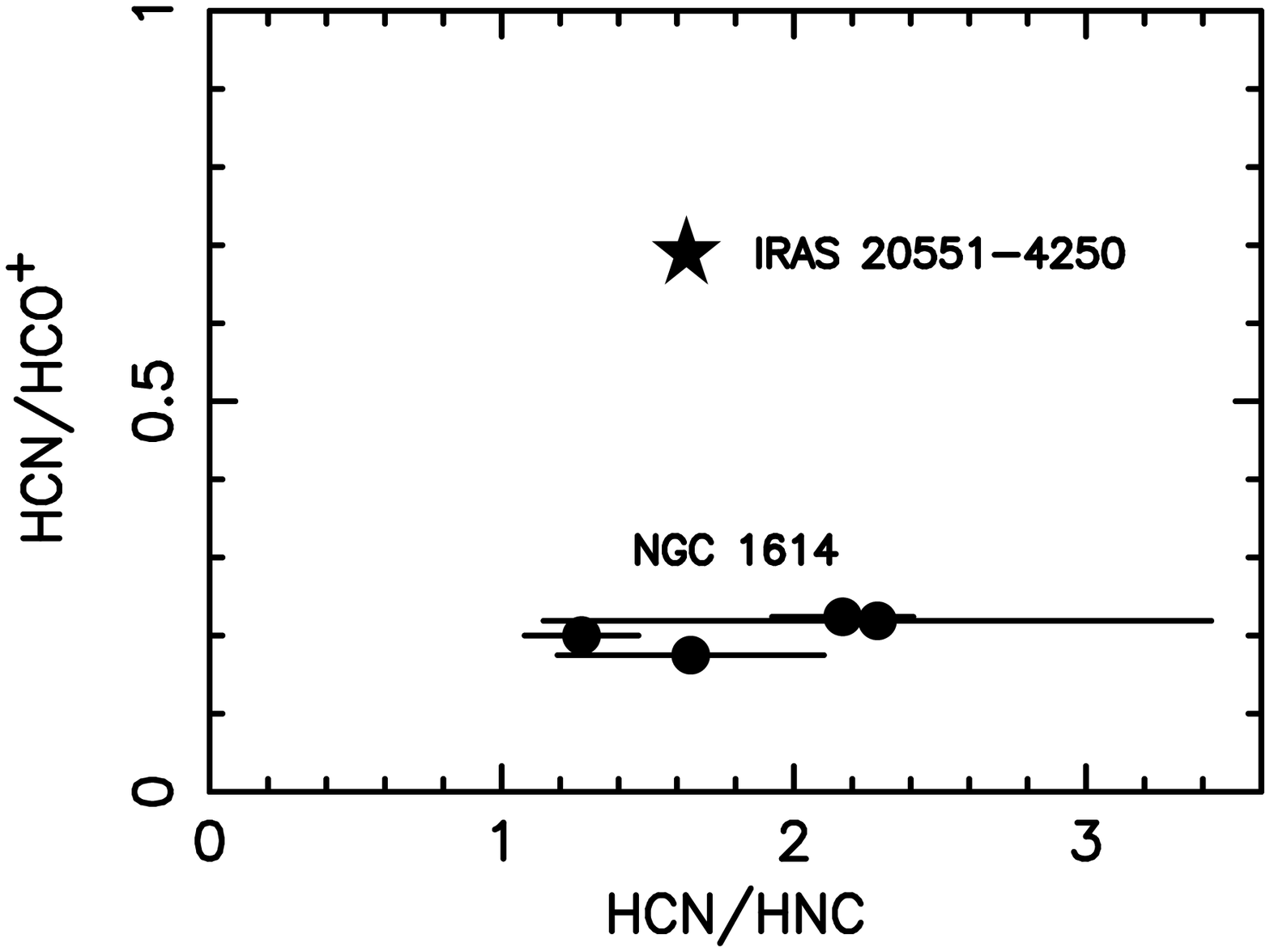} 
\end{center}
\caption{
HCN-to-HNC (abscissa) and HCN-to-HCO$^{+}$ (ordinate) flux 
ratios at J=4--3.
The filled star marks the ratio for IRAS 20551$-$4250.
The filled circles are four data points for the starburst galaxy NGC
1614 \citep{ima13}.    
}
\end{figure}


\begin{thebibliography}{}
\bibitem[Alonso-Herrero et al.(2009)]{alo09}
         Alonso-Herrero, A., Garcia-Marin, M., Monreal-Ibero, A.,
         Colina, L., Arribas, S., Alfonso-Garzon, J., \&  Labiano,
         A. 2009, A\&A, 506, 1541 
\bibitem[Aalto et al.(1995)]{aal95}
         Aalto, S., Booth, R. S., Black, J. H., \& Johansson,
         L. E. B. 1995, A\&A, 300, 369 
\bibitem[Aladro et al.(2011)]{ala11}
         Aladro, R., Martin, S., Martin-Pintado, J., Mauersberger, R.,
         Henkel, C., Ocana Flaquer, B., Amo-Baladron, M. A. 2011, A\&A,
         535, 84
\bibitem[Aladro et al.(2013)]{ala13}
         Aladro, R., et al. 2013, A\&A, 549, 39
\bibitem[Caputi et al.(2007)]{cap07} 
         Caputi, K. I., et al. 2007, ApJ, 660, 97
\bibitem[Costagliola et al.(2011)]{cos11}
         Costagliola, F., et al. 2011, A\&A, 528, 30
\bibitem[Duc et al.(1997)]{duc97}
         Duc, P. -A., Mirabel, I. F., \& Maza, J. 1997, A\&AS, 124, 533 
\bibitem[Evans et al.(2003)]{eva03} 
         Evans, A. S. et al. 2003, AJ, 125, 2341
\bibitem[Franceschini et al.(2003)]{fra03}
         Franceschini, A. et al. 2003, MNRAS, 343, 1181  
\bibitem[Garcia-Burillo et al.(2006)]{gar06}
         Garcia-Burillo, S., et al. 2006, ApJ, 645, L17
\bibitem[Garcia-Burillo et al.(2010)]{gar10}
         Garcia-Burillo, S., et al. 2010, A\&A, 519, 2
\bibitem[Goto et al.(2010)]{got10}
         Goto, T., et al. 2010, A\&A, 514, 6 
\bibitem[Gracia-Carpio et al.(2006)]{gra06}
         Gracia-Carpio, J., Garcia-Burillo, S., Planesas, P., \& Colina, L.
          2006, ApJ, 640, L135
\bibitem[Harada et al.(2010)]{har10}
         Harada, N., Herbst, E., \& Wakelam, V. 2010, ApJ, 721, 1570 
\bibitem[Hopkins et al.(2006)]{hop06}
         Hopkins, P. F., Hernquist, L., Cox, T. J., Di Matteo, T.,
         Robertson, B., \& Springel, V. 2006, ApJS, 163, 1
\bibitem[Imanishi et al.(2011)]{ima11} 
         Imanishi, M., Imase, K., Oi, N., \& Ichikawa, K. 2011, AJ, 141,
         156
\bibitem[Imanishi et al.(2010a)]{ima10}
         Imanishi, M., Nakagawa, T., Shirahata, M., Ohyama, Y., \&
         Onaka, T. 2010a, ApJ, 721, 1233
\bibitem[Imanishi \& Nakanishi(2006)]{in06} 
         Imanishi, M., \& Nakanishi, K. 2006, PASJ, 58, 813
\bibitem[Imanishi \& Nakanishi(2013)]{ima13} 
         Imanishi, M., \& Nakanishi, K. 2013, AJ, 146, 47
\bibitem[Imanishi et al.(2004)]{ima04}
         Imanishi, M., Nakanishi, K., Kuno, N., \& Kohno, K. 2004, AJ,
         128, 2037 
\bibitem[Imanishi et al.(2006)]{ima06} 
         Imanishi, M., Nakanishi, K., \& Kohno, K. 2006, AJ, 131, 2888
\bibitem[Imanishi et al.(2007)]{ima07} 
         Imanishi, M., Nakanishi, K., Tamura, Y., Oi, N., \& Kohno,
         K. 2007, AJ, 134, 2366
\bibitem[Imanishi et al.(2009)]{ima09} 
         Imanishi, M., Nakanishi, K., Tamura, Y., \& Peng, C. -H. 2009,
         AJ, 137, 3581 
\bibitem[Imanishi et al.(2010b)]{ima10b} 
         Imanishi, M., Nakanishi, K., Yamada, M., Tamura, Y., \& Kohno, K. 
         2010b, PASJ, 62, 201
\bibitem[Kohno(2005)]{koh05}
         Kohno, K. 2005, in AIP Conf. Ser. 783, 
         The Evolution of Starbursts, ed. S. H\"uttemeister, E. Manthey,
         D. Bomans, \& K. Weis (New York: AIP), 203 (astro-ph/0508420)
\bibitem[Komatsu et al.(2009)]{kom09}
         Komatsu, E., et al. 2009, ApJS, 180, 330
\bibitem[Krips et al.(2008)]{kri08}
         Krips, M., Neri, R., Garcia-Burillo, S., Martin, S., Combes,
         F., Gracia-Carpio, J., \& Eckart, A. 2008, ApJ, 677, 262 
\bibitem[Lahuis et al.(2007)]{lah07}
         Lahuis, F. et al. 2007, ApJ, 659, 296 
\bibitem[Lintott \& Viti(2006)]{lin06}
         Lintott, C., \& Viti, S. 2006, ApJ, 646, L37
\bibitem[Magnelli et al.(2011)]{mag11}
         Magnelli, B., Elbaz, D., Chary, R. R., Dickinson, M., Le
         Borgne, D., Frayer, D. T., \& Willmer, C. N. A. 2011, A\&A,
         528, 35 
\bibitem[Martin et al.(2006)]{mar06}
         Martin, S., Mauersberger, R., Martin-Pintado, J., Henkel, C.,
         Garcia-Burillo, S. 2006, ApJS, 164, 450   
\bibitem[Meijerink et al.(2007)]{mei07}
         Meijerink, R., Spaans, M., \& Israel, F. P. 2007, A\&A, 461, 793
\bibitem[Murphy et al.(2011)]{mur11}
         Murphy, E. J., Chary, R. -R., Dickinson, M., Pope, A., Frayer,
         D. T., \& Lin, L. 2011, ApJ, 732, 126
\bibitem[Nakanishi et al.(2005)]{nak05}
         Nakanishi, K., Okumura, S. K., Kohno, K., Kawabe, R., \&
         Nakagawa, T. 2005, PASJ, 57, 575
\bibitem[Nardini et al.(2010)]{nar10}
         Nardini, E., Risaliti, G., Watabe, Y., Salvati, M., \& Sani,
         E. 2010, MNRAS, 405, 2505
\bibitem[Perez-Beaupuits et al.(2007)]{per07}
         Perez-Beaupuits, J. P., Aalto, S., \& Gerebro, H. 2007, A\&A,
         476, 177 
\bibitem[Rangwala et al.(2011)]{ran11}
         Rangwala, N., et al. 2011, ApJ, 743, 94 
\bibitem[Risaliti et al.(2006)]{ris06}
         Risaliti, G., et al. 2006, MNRAS, 365, 303
\bibitem[Sakamoto et al.(2010)]{sak10}
         Sakamoto, K., Aalto, S., Evans, A. S., Wiedner, M., \& Wilner,
         D. 2010, ApJ, 725, L228
\bibitem[Sakamoto et al.(2013)]{sak13}
         Sakamoto, K., Aalto, S., Costagliola, F., Martin, S., Ohyama,
         Y., Wiedner, M. C., \& Wilner, D. J. 2013, ApJ, 764, 42 
\bibitem[Sanders \& Ishida(2004)]{san04}
         Sanders, D. B., \& Ishida, C. M. 2004, ASPC, 320, 230
\bibitem[Sanders \& Mirabel(1996)]{sam96}
         Sanders, D. B., \& Mirabel, I. F. 1996, ARA\&A, 34, 749
\bibitem[Sani et al.(2008)]{san08}
         Sani, E., et al. 2008, ApJ, 675, 96  
\bibitem[Schilke et al.(1997)]{sch97}
         Schilke, P., Groesbeck, T. D., Blake, G. A., \& Phillips,
         T. G. 1997, ApJS, 108, 301
\bibitem[Spoon et al.(2001)]{spo01} 
         Spoon, H. W. W., Keane, J. V., Tielens, A. G. G. M., Lutz, D.,
         \& Moorwood, A. F. M. 2001, A\&A, 365, L353 
\bibitem[Sutton et al.(2001)]{sut91}
         Sutton, E. C., Jaminet, P. A., Danchi, W. C., \& Blake, G. A.
         1991, ApJS, 77, 255 
\bibitem[Tacconi et al.(1994)]{tac94}
         Tacconi, L.  J., Genzel, R., Blietz, M., Cameron, M., Harris, A. I., 
         \& Madden, S. 1994, ApJ, 426, L77
\bibitem[Weiss et al.(2007)]{wei07}
        Weiss, A., et al.  2007, A\&A, 467, 955
\end{thebibliography}
\end{document}